\documentclass[twocolumn]{aastex7}

\usepackage{enumitem}
\usepackage{multirow}

\begin{document}

\newcommand{\nh}{$N_H$}
\newcommand{\source}{MAXI J1807$+$132}
\newcommand{\nicer}{\textit{NICER}}
\newcommand{\swift}{\textit{Swift}}
\newcommand{\dbb}{\texttt{diskbb}}
\newcommand{\bb}{\texttt{blackbody}}
\newcommand{\nthc}{\texttt{nthComp}}
\newcommand{\tbabs}{\texttt{TBabs}}
\newcommand{\tbfeo}{\texttt{TBfeo}}
\newcommand{\tbvarabs}{\texttt{TBvarabs}}
\newcommand{\chisq}{$\chi^{2}$}
\newcommand{\kte}{$kT_e$}
\newcommand{\inptype}{$inp{type}$}
\newcommand{\tin}{$T_{in}$}
\newcommand{\kt}{$kT$}
\newcommand{\ktbb}{$kT_{bb}$}
\newcommand{\gama}{$\Gamma$}
\newcommand{\norm}{\texttt{norm}}
\newcommand{\gp}{$g^\prime$}
\newcommand{\rp}{$r^\prime$}
\newcommand{\ip}{$i^\prime$}
\newcommand{\zs}{$z_{s}$}
\newcommand{\uwt}{$W2$}
\newcommand{\uvv}{$V$}
\newcommand{\nub}{$\nu_{b}$}

\graphicspath{{./}{./figures/}}

\title{A Multi-wavelength Characterization of the 2023 Outburst of MAXI J1807$+$132: Manifestations of Disk Instability and Jet Emission.}

\author[0000-0001-7590-5099]{Sandeep K. Rout}
\affiliation{New York University Abu Dhabi, PO Box 129188, Abu Dhabi, UAE}
\affiliation{Center for Astrophysics and Space Science (CASS), New York University Abu Dhabi, PO Box 129188, Abu Dhabi, UAE}
\email[show]{sandeep.rout@nyu.edu}

\author[0000-0003-1285-4057]{M. Cristina Baglio}
\affiliation{INAF--Osservatorio Astronomico di Brera, Via Bianchi 46, I-23807 Merate (LC), Italy}
\email[]{maria.baglio@inaf.it}

\author[0000-0003-0764-0687]{Andrew Hughes}
\affiliation{Department of Physics, University of Oxford, Denys Wilkinson Building, Keble Road, Oxford OX1 3RH, UK}
\affiliation{Department of Physics, University of Alberta, Edmonton, T6G 2E1, Canada}
\email[]{hughes1@ualberta.ca}

\author[0000-0002-3500-631X]{David M. Russell}
\email[]{dr124@nyu.edu}
\affiliation{New York University Abu Dhabi, PO Box 129188, Abu Dhabi, UAE}
\affiliation{Center for Astrophysics and Space Science (CASS), New York University Abu Dhabi, PO Box 129188, Abu Dhabi, UAE}

\author[0000-0002-1583-6519]{D. M. Bramich}
\email[]{}
\affiliation{New York University Abu Dhabi, PO Box 129188, Abu Dhabi, UAE}
\affiliation{Center for Astrophysics and Space Science (CASS), New York University Abu Dhabi, PO Box 129188, Abu Dhabi, UAE}
\email[]{dmb20@nyu.edu}

\author[0000-0002-5319-6620]{Payaswini Saikia}
\email[]{ps164@nyu.edu}
\affiliation{New York University Abu Dhabi, PO Box 129188, Abu Dhabi, UAE}
\affiliation{Center for Astrophysics and Space Science (CASS), New York University Abu Dhabi, PO Box 129188, Abu Dhabi, UAE}

\author[0000-0003-0168-9906]{Kevin Alabarta}
\email[]{kalabarta@nyu.edu}
\affiliation{New York University Abu Dhabi, PO Box 129188, Abu Dhabi, UAE}
\affiliation{Center for Astrophysics and Space Science (CASS), New York University Abu Dhabi, PO Box 129188, Abu Dhabi, UAE}

\author[0000-0002-4344-7334]{Montserrat Armas Padilla}
\affiliation{Instituto de Astrof\'{i}sica de Canarias (IAC), V\'{i}a L\'{a}ctea s/n, La Laguna E-38205, S/C de Tenerife, Spain}
\affiliation{Departamento de Astrof\'{i}sica, Universidad de La Laguna, La Laguna E-38205, S/C de Tenerife, Spain}
\email[]{m.armaspadilla@gmail.com}

\author[0000-0001-6278-1576]{Sergio Campana}
\affiliation{INAF--Osservatorio Astronomico di Brera, Via Bianchi 46, I-23807 Merate (LC), Italy}
\email[]{sergio.campana@inaf.it}

\author[0000-0001-9078-5507]{Stefano Covino}
\affiliation{INAF--Osservatorio Astronomico di Brera, Via Bianchi 46, I-23807 Merate (LC), Italy}
\affiliation{Como Lake centre for AstroPhysics (CLAP), DiSAT, Universit\'a dell’Insubria, via Valleggio 11, 22100 Como, Italy}
\email[]{stefano.covino@brera.inaf.it}

\author[0000-0001-7164-1508]{Paolo D$^\prime$Avanzo}
\affiliation{INAF--Osservatorio Astronomico di Brera, Via Bianchi 46, I-23807 Merate (LC), Italy}
\email[]{paolo.davanzo@inaf.it}

\author[0000-0002-5654-2744]{Rob Fender}
\affiliation{Department of Physics, University of Oxford, Denys Wilkinson Building, Keble Road, Oxford OX1 3RH, UK}
\email[]{rob.fender@physics.ox.ac.uk}

\author[0000-0001-5638-5817]{Paolo Goldoni}
\affiliation{Universit\'{e} Paris Cit\'{e}, CNRS, CEA, Astroparticule et Cosmologie, F-75013 Paris, France}
\email[]{goldoni@apc.univ-paris7.fr}

\author[0000-0001-8371-2713]{Jeroen Homan}
\affiliation{Eureka Scientific, Inc., 2452 Delmer Street, Oakland, CA 94602, USA}
\email[]{jeroenhoman@icloud.com}

\author{Fraser Lewis}
\affiliation{Faulkes Telescope Project, School of Physics and Astronomy, Cardiff University, The Parade, Cardiff, CF24 3AA Wales, UK}
\affiliation{The Schools’ Observatory, Astrophysics Research Institute, Liverpool John Moores University, 146 Brownlow Hill, Liverpool L3 5RF, UK}
\email[]{fraser.lewis68@gmail.com}

\author[0000-0001-9487-7740]{Nicola Masetti}
\affiliation{INAF--Osservatorio di Astrofisica e Scienza dello Spazio, Via Piero Gobetti 101, I-40129 Bologna, Italy}
\affiliation{Universidad Andr\'es Bello, Av. Fern\'andez Concha 700, Las Condes, Santiago, Chile}
\email[]{nicola.masetti@inaf.it}

\author[0000-0002-6154-5843]{Sara Motta}
\affiliation{INAF--Osservatorio Astronomico di Brera, Via Bianchi 46, I-23807 Merate (LC), Italy}
\email[]{sara.motta@inaf.it}

\author[0000-0002-3348-4035]{Teo Mu\~{n}oz-Darias}
\affiliation{Instituto de Astrof\'{i}sica de Canarias (IAC), V\'{i}a L\'{a}ctea s/n, La Laguna E-38205, S/C de Tenerife, Spain}
\affiliation{Departamento de Astrof\'{i}sica, Universidad de La Laguna, La Laguna E-38205, S/C de Tenerife, Spain}
\email[]{tmd.astronomy@gmail.com}

\author[0000-0001-6289-7413]{Alessandro Papitto}
\affiliation{INAF Osservatorio Astronomico di Roma, Via Frascati 33, I-00078 Monte Porzio Catone (RM), Italy}
\email[]{alessandro.papitto@inaf.it}

\author[0000-0002-7930-2276]{Thomas D. Russell}
\affiliation{INAF, Istituto di Astrofisica Spaziale e Fisica Cosmica, Via U. La Malfa 153, I-90146 Palermo, Italy}
\email[]{thomas.russell@inaf.it}

\author{Gregory Sivakoff}
\affiliation{Department of Physics, University of Alberta, Edmonton, T6G 2E1, Canada}
\email[]{sivakoff@ualberta.ca}

\author[0000-0002-5686-0611]{Jakob van den Eijnden}
\affiliation{Anton Pannekoek Institute for Astronomy, Universiteit van Amsterdam, Science Park 904, 1098, XH, Amsterdam, The Netherlands}
\email[]{a.j.vandeneijnden@uva.nl}

\correspondingauthor{Sandeep K. Rout}

\begin{abstract}

Several phenomenological aspects of low-luminosity neutron star transients, such as atolls, remain poorly understood. One such source, \source{}, entered its latest outburst in July 2023. To thoroughly characterize this outburst, we conducted an extensive observational campaign spanning radio to X-ray wavelengths. Here, we present the results of this campaign, which covered the period from before the outburst to the return to quiescence. We detected a delay between the X-ray and optical rise times, which is consistent with the predictions of the disk instability model with a truncated disk. The color evolution and optical/X-ray correlations, along with infrared and radio detections, support the presence of jet synchrotron emission during the gradual decay phase following the peak. We also report for the first time in an X-ray binary a near-orthogonal rotation of the optical polarization just before a small flare, after which the jet is thought to be quenched. The main outburst is followed by several high-amplitude, rapid reflares in the optical, ultraviolet, and X-ray bands, the origin of which remains difficult to constrain.

\end{abstract}


\section{Introduction} \label{sec:intro}

Low-mass X-ray binaries (LMXBs) are transient systems in which a compact object, i.e., a black hole (BH) or neutron star (NS), accretes matter from a low-mass star via Roche-lobe overflow. The outburst and quiescence cycle in LMXBs is typically explained by the disk-instability model \citep[DIM; see][for reviews]{dubus01, lasota01}. Although DIM was originally developed for dwarf novae \citep{smak71}, it can also be applied to NS and BH LMXBs due to their morphological similarities and the similarities in the rapid rise and exponential decay behavior observed in both types of system \citep{vanparadijs84, vanparadijs96}. To explain the typical phenomenology of LMXBs, several modifications to the DIM have been proposed over the years. Notable modifications have been the inclusion of irradiation effects on the outer disk and companion star, the impact of outflows such as winds and jets on the disk structure, inner disk truncation, and fluctuations in the mass transfer rate \citep[e.g.,][]{hameury20}.

The accretion process is often accompanied by the ejection of matter from the system in the form of winds and jets \citep[e.g.,][]{fender14, fender16, homan16}. These outflows play a crucial role in the outward transfer of angular momentum and in enriching the surrounding medium with stellar material. Jets often appear as relativistic outflows along the spin axis of the compact object, which are at times found to be misaligned with the orbital plane and undergo precession. The strong magnetic fields that channel these outflows give rise to synchrotron emission, which typically follows a broken power-law spectrum \citep[e.g.,][]{blandford79, falcke04}. The frequency at which the break occurs, which typically falls in the mid infrared to optical frequency range, is an important observable for determining the total energy output of the jet \citep[e.g.,][]{russellt14}. However, measuring the break frequency is not straightforward, as it requires simultaneous broadband coverage spanning several orders of magnitude in frequency \citep[e.g.,][]{russelld13a}.    

In BH LMXBs, jet emission is strongly linked to the spectral state of the source. Compact jets are typically detected in the low-hard state, while ballistic jets are observed during the transition from the hard-intermediate to the soft-intermediate states \citep{fender04b, fender09, carotenuto21}. NS systems are sub-classified into Z and atoll sources based on their position on the X-ray color-color diagram \citep[][]{hasinger89}. They mainly differ in their luminosities, with the former being brighter than the latter. In Z sources, jets are brightest in the horizontal branch and fade along the normal and flaring branches \citep{penninx88, hjellming90, migliari06a}. Similarly, in atoll sources, jets are typically brighter in the hard island branches and fainter in the banana branches. In general, jets are brighter in the low accretion rate, radiatively inefficient phases, and fainter in high accretion rate states. However, unlike their BH counterparts, the association between jet brightness and spectral states is not as strict in NS sources. For example, an increase in the radio flux was detected when the Z source GX $17+2$ moved from the horizontal to the normal branch \citep{tan92}. Similarly, \citet{migliari04} detected radio emission in the soft banana branches of the atoll sources 4U $1820-30$ and Ser X-1 \citep[see also][for a detailed discussion on 4U 1820$-30$]{russellt21}. Furthermore, the jets in the NS systems are approximately 22 times fainter than the BH jets, making their study more challenging \citep[e.g.,][]{fender01b, russelld07, tudor17, gallo18}. Even the radio/X-ray correlation slopes of Aql X-1, a well-studied atoll source, are complex and debated \citep[e.g.,][]{tudose09, gusinskaia20,fijma23}. Although the radio luminosity of the NS jets is not fully understood, they are generally still correlated with the X-ray emission to some extent \citep[including in the source that switches between Z and atoll states, XTE J1701--462;][]{gasealahwe24}.

With the aim of expanding our understanding of the accretion-ejection paradigm in NS LMXBs, we conducted an extensive multi-wavelength observing campaign on the atoll source \source{} during its latest outburst in July 2023. \source{} had undergone two previous outbursts in 2017 and 2019, during which its NS nature was confirmed by the detection of Type-I thermonuclear bursts \citep{shidatsu17, jiminezibarra19, albayati21}. We reported the X-ray properties of the source with data from the Neutron Star Interior Composition Explorer (\nicer{}) in a separate paper \citep{rout25}. In this work, we extend the study of the properties of the source to the rest of the wavelengths, i.e., ultraviolet (UV) to radio. We rely on the spectral state classification from the X-ray study and also use the $2-10$ keV flux values derived from it to study the optical/X-ray correlations. We provide a summary of all the observations used in this work in Table \ref{tab:obslogs}. In Section \ref{sec:obs} we give details of all the data used in the work along with their reduction. The different diagnostic tools and the main results are detailed in Section \ref{sec:analysis}. In Section \ref{sec:discussion} we provide an interpretation of the results and compare them with the literature. Finally, in Section \ref{sec:summary} we summarize the main results of this work.

\begin{table*}[]

\caption{Summary of all observations used in this work. The last column contains the total number of observations in all the filters/bands mentioned in the third column.}
\label{tab:obslogs}
\centering                       
\begin{tabular}{ccccc}

\hline

Bands & Instruments & Energy/Wavelength & MJD & No. of Observations \\
\hline \hline
\multirow{2}{3em}{X-ray} & \nicer{}/XTI & $0.5-10$ keV & 60132 - 60161 & 38 \\
 & \swift{}/XRT & $0.7-10$ keV & 60137 - 60238 & 23\\
\hline
Ultraviolet & \swift{}/UVOT & \uwt{} (1928 \AA), $M2$ (2246 \AA), $W1$ (2600 \AA) &  60137 - 60231 & 16 \\
\hline
\multirow{3}{3em}{Optical} & \swift{}/UVOT & $U$ (3465 \AA), $V$ (4392 \AA) &  60137 - 60228 & 18 \\

& Faukes, LCO & \gp{} (4770 \AA), \rp{} (6215 \AA), \ip{} (7545 \AA), \zs{} (8700 \AA)  & 59326 - 60239 & 241  \\

& VLT/FORS2  & $B$ (440 nm), $V$ (557 nm), $R$ (655 nm), $I$ (768 nm)  & 60142, 60143, 60149 & 3 \\
\hline
Infrared & REM & $H$ (1.64 $\mu$m) & 60138 - 60184 & 7  \\
\hline
\multirow{2}{3em}{Radio} & VLA & A-band (1.5 GHz), C-band (6 GHz) & 60202 - 60226 & 7 \\

& MeerKAT  & L-band (1.28 GHz) & 60132 - 60231 & 15 \\

\hline
\end{tabular}

\end{table*}

\section{Observations and Data Reduction} \label{sec:obs}

\subsection{X-ray with \nicer{} and \swift{}}

We used data from both \swift{} and \nicer{} to study the X-ray emission from the source. The analysis of \nicer{} \citep{gendreau16} data was carried out using \texttt{NICERDAS\_V011a}. We followed the standard procedures for the extraction of the lightcurves and spectra using the \texttt{nicerl3} tools. More details on the reduction are given in \citet{rout25}. We used the \nicer{} data to study two things: 1) the delay between the X-ray and optical rise, and 2) correlations between the optical and X-ray emission. Specifically for the correlations, we used the unabsorbed flux in the $2-10$ keV band from the spectral fits \citep{rout25}. The lightcurves and spectra for \swift{}/X-ray Telescope \citep[XRT;][]{burrows05} were extracted from the XRT data products generator\footnote{\url{https://www.sw 1ift.ac.uk/user_objects/docs.php}} \citep{evans07,evans09}. This corrects for pile-up effects by selecting an annular extraction region with appropriate inner radius depending on the count rate. Similarly, the effects of grade migration in the piled-up data acquired in the photon-counting (PC) mode are mitigated by selecting grade $0-4$ events.

\subsection{Optical and Ultraviolet with \swift{}}

\swift{} started observations of \source{} as soon as the source reached the peak of the 2023 outburst. Both XRT and the Ultraviolet and Optical Telescope \citep[UVOT;][]{roming05} continued the monitoring campaign for about three months after the main outburst and captured several reflaring events. UV photometry was done using the \texttt{uvotsource} module in \texttt{heasoft-6.32}. A $5^{''}$ circular region centered at the source coordinates was selected for the source photometry and a $\sim 10^{''}$ circle offset from the source was chosen for the background. Photometric measurements were carried out for the $W1$, $M2$, $W2$, $U$, and \uvv{} bands. Some observations were also made in the $B$ band. However, they were acquired near the quiescence level and the source was not significantly detected. Hence, we did not include them in our analysis. Of all of these filters, observations with the \uvv{} and $W2$ bands were made at simultaneous epochs and with good cadence during the main outburst. Hence, they are used for studying the color evolution and spectral energy distributions (SEDs).

\subsection{Optical photometry with LCO}

\source{} is a regular target in our LMXB monitoring program with the 2m Faulkes Telescope and the 1m telescopes of the Las Cumbres Observatory (LCO) network \citep{lewis08a}. With our fully automated pipeline X-ray Binary New Early Warning System \citep[XB-NEWS;][]{russelld19, goodwin20}, we are able to create near real-time lightcurves of the source in the optical wavebands. As a result of this, we detected the optical rise on 2nd July 2023, and a multi-wavelength observation campaign was started after that \citep{atelsaikia23a, atelilliano23, atelsaikia23b}. 

The optical photometry was carried out using the XB-NEWS pipeline for the SDSS filters \zs{}, \ip{}, \rp{}, and \gp{} (Alabarta et al. 2025, to be submitted). The pipeline scans the LCO archive at regular intervals and,  whenever new observations are available, it downloads the reduced images. Poor quality images are immediately rejected. For those images that pass the quality control stage, an astrometric solution is derived by comparison to the \textit{Gaia} DR2 catalog\footnote{\url{https://www.cosmos.esa.int/web/gaia/dr2}}, and then multi-aperture photometry \citep[MAP;][]{stetson90} is performed for all detected sources. Flux calibration is done using an enhanced version of  the ATLAS-REFCAT2 catalog\footnote{\url{https://archive.stsci.edu/prepds/atlas-refcat2/}}, which includes other catalogs such as Pan-STARRS DR1\footnote{\url{https://panstarrs.stsci.edu}}, APASS\footnote{\url{https://www.aavso.org/apass}}, etc. \citep{tonry18}, by solving for zero-point offsets along the lines of the method given in \citet{bramich12}. When the main target is not formally detected due to faintness, forced photometry is carried out at the target coordinates. Magnitude measurements of the target with uncertainties that exceed 0.25 mag are rejected.

\subsection{Optical polarimetry with the VLT}

We observed MAXI J1807$+$132 with the Very Large Telescope (VLT), at Paranal Observatory (Chile), equipped with the FOcal Reducer/low dispersion Spectrograph 2 \citep[FORS2;][]{appenzeller98} instrument in polarimetric mode using the following set of four optical filters: $b\_HIGH+113$ ($B$; central wavelength 440 nm), $v\_HIGH+114$ (\uvv{}; central wavelength 557 nm), $R\_SPECIAL +76$ ($R$; central wavelength 655 nm), $I\_BESS+77$ ($I$; central wavelength 768 nm). Observations were made on three epochs during the outburst: 2023 Jul 17 (MJD 60142.1), 2023 Jul 18 (MJD 60143.1), and 2023 Jul 24 (MJD 60149.0). 

A Wollaston prism was inserted into the instrument's optical path, splitting the incoming light into two beams with orthogonal polarization (ordinary and extraordinary). To prevent these beams from overlapping on the CCD, a mask was employed. Additionally, a rotating half-wave plate (HWP) was incorporated, enabling the capture of images at four distinct angles ($\Phi_i$) relative to the telescope position angle: $\Phi_{i}=22.5^{\circ}(i-1)$ with $i=1, 2, 3, 4$. 

With this setup, one image for each of the four HWP angles was captured in each filter for each night employing the following exposure times for each image: 65 s in $B$, 45 s in \uvv{}, 40 s in $R$ and 70 s in $I$. All images were then processed by subtracting an averaged bias frame and normalizing with a flat field. Aperture photometry was performed using \texttt{daophot} \citep{Stetson1987}, applying a 6-pixel radius aperture (which corresponds to $\sim 0.8''$).

The linear polarization degree $P$ and polarization angle $\theta$ are calculated following the methodology detailed in Section 3.1.1 of \citet{baglio20}. These parameters are not corrected for any instrumental contributions to linear polarization. This is justified because unpolarized standard stars routinely observed with FORS2 show that the instrument's polarization has remained stable and very low (below $0.3\%$, on-axis) across all bands over the past decade. More details on the calculation of the linear polarization and estimation of the posterior probabilities are given in Section \ref{sec:optpol}.

\subsection{Near-Infrared with REM}

We observed MAXI J1807+132 in the $H$ band with the 60-cm Robotic Eye Mount (REM) telescope at La Silla (Chile) from July 13 (MJD 60138) to August 28, 2023 (MJD 60184), for a total of 7 epochs of observations randomly distributed within this date range. For each epoch, three sets of observations were taken, except for MJD 60138 and MJD 60150 where only one set of observations was acquired. Each set consisted of five dithered 30-second integration exposures that were combined for optimal background subtraction. Unfortunately, the signal-to-noise ratio was not always good enough to detect the target, typically due to bad seeing conditions (which reached up to $5''$). In the images where detection was possible, we performed aperture photometry using {\tt daophot} \citep{Stetson1987}, applying an aperture with radius of $2$ times the average full width at half maximum (FWHM) of the flux profiles of the field stars. In the cases where a detection was not possible, we estimated 3$\sigma$ upper limits to the target flux for those images. Flux calibration was performed against the 2MASS\footnote{https://irsa.ipac.caltech.edu/Missions/2mass.html} catalog \citep{skrutskie06}.

\subsection{Radio with MeerKAT}

We observed MAXI J1807$+$132 with MeerKAT \citep{Jonas16_AKH} as part of the large survey project ThunderKAT \citep{tkat17_AKH}. We began our monitoring on 7 July 2023 (MJD 60132), continuing until 14 October 2023 (MJD 60231), totaling 15 observations. Each observation consisted of a single scan of 15 minutes on-source flanked by two 2-minute scans of a nearby gain calibrator (J1733$-$1304). Each epoch also included a 5-minute PKS B1934-638 (J1939$-$6342) scan for flux and bandpass calibration. All MeerKAT observations used the L-band receiver, with a central frequency of 1.28\,GHz and a $\sim$\,856 MHz bandwidth. We processed the MeerKAT data with the semi-automated pipeline \textsc{oxkat} \citep{oxkat20_AKH}, producing a single image per observing epoch. A comprehensive description of \textsc{oxkat} is found in \citet{mightee20_AKH}. The morphology of the radio emission from MAXI J1807$+$132 showed temporal variability. To verify its association with the source, we obtained observations with the Very Large Array (VLA). Details on the VLA campaign and its analysis are given in Appendix \ref{sec:vla}.

\section{Analysis and Results} \label{sec:analysis}

\subsection{Multi-wavelength lightcurve and X-ray delay} \label{sec:ana-delay}

\begin{figure*}
    \centering
    \includegraphics[scale=.7]{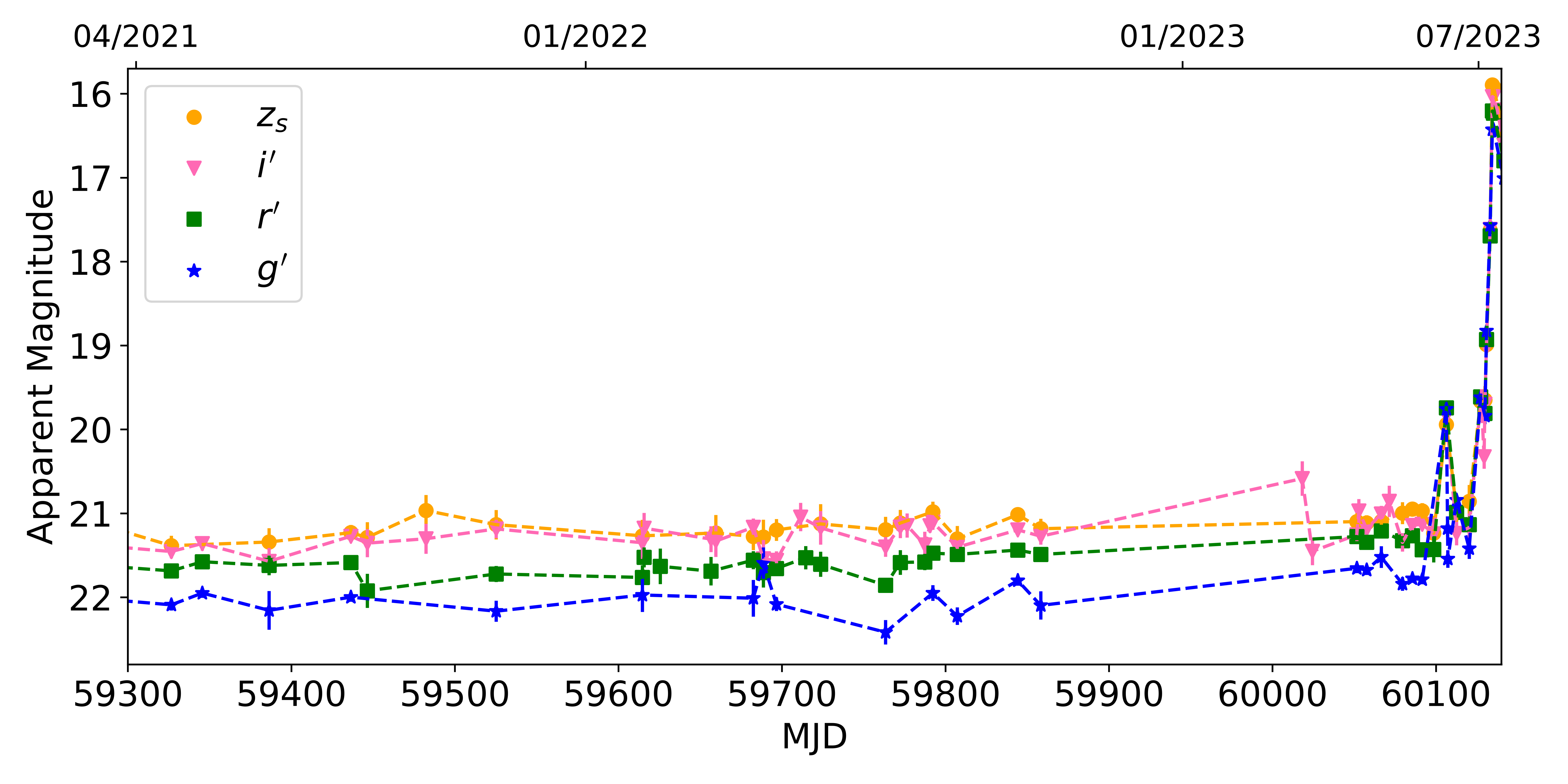}
    \caption{The quiescent optical lightcurve of \source{} since April 2021, when our regular LCO monitoring started with the SDSS filters, until the beginning of the latest outburst in July 2023.}
    \label{fig:lcquiesc}
\end{figure*}

The long-term optical lightcurve of \source{} since April 2021 (MJD 59326) until the beginning of the 2023 outburst is shown in Figure \ref{fig:lcquiesc}. The source remained dormant for nearly 20 months until November 2022 (MJD 59894). Following this, there was a four-month gap in our observations. When the observations resumed, the fluxes in the \gp{} and \rp{} bands, and less so in the \ip{} band, were at a slightly elevated level (by $\sim 0.1-0.2$ mag). To determine if there is a significant difference in the quiescence brightness level before and after the four-month observational gap, we did the following. First, we fitted all of the data points between MJD 59300 and 60100 with a constant-brightness model (M1) using $\chi^2$-minimization. Then we fitted a second model to the same data, again using $\chi^2$-minimization, where this model has different constant brightnesses for before and after the observational gap (M2). The $\Delta \chi^2$ values between the two models are 58.62, 117.84, 12.78, and 9.00 for the \gp{}, \rp{}, \ip{}, and \zs{} bands, respectively. Given that there is one more free parameter in M2 compared to M1, we can compare these values with a $\chi^2$ distribution with one degree of freedom. In doing so, we find that the $\Delta \chi^2$ values are significant at the $1~\%$ level (i.e., they are greater than the threshold of $\sim6.62$), thereby leading us to reject M1 at this level of significance. LMXBs are sometimes known to undergo a slow rise phase a few years before the onset of the outburst \citep[e.g.,][and references therein]{russelld18b}. We tested this possibility by fitting the long-term lightcurve with a model consisting of a constant magnitude up to a certain point in time (break point) followed by a linear trend (M3). We iterated the fits for a grid of 8000 break points between MJD 59300 and MJD 60100, and computed the $\chi^2$ for each fit. The best fits for the \gp{} and \rp{} bands occur at the break points of MJD $59763.4^{+27.5}_{-58.8}$ and MJD $59696.6^{+39.8}_{-55.3}$, respectively. However, for the \ip{} and \zs{} bands, the best-fit break points were found at the lower limit of the time range and therefore could not be constrained. The $90\%$ upper limits of the break points for the \ip{} and \zs{} bands are MJD 59652.9 and MJD 59743.1, respectively. This model provided a much better fit compared to M1 with $\Delta \chi^2$ between the two models being 60.10, 129.61, 19.67, and 14.13 for the \gp{}, \rp{}, \ip{}, and \zs{} bands, respectively (again significant at the 1$\%$ level). However, M2 is not nested within M3, and both have exactly the same number of free parameters. Hence, we computed the likelihood ratios M3 to M2 for the four filters, which are 2.1, 358.7, 31.3, and 16.3. Hence, the linear rise is significantly favored over a sudden increase in brightness in three out of four filters. We conclude that the source indeed shows a slow rise $\sim400$ days before the main outburst, with some hints of the redder band fluxes rising earlier than the bluer band fluxes. The rise rates in the \gp{} ($0.34\pm0.07$ mag/year) and \rp{} ($0.30\pm0.03$ mag/year) bands are steeper than in the \ip{} ($0.11\pm0.03$ mag/year) and \zs{} ($0.14\pm0.03$ mag/year) bands. We show the fits of M3 to the long-term lightcurves in Figure \ref{fig:slowrise}.

\begin{figure}
    \centering
    \includegraphics[scale=0.5]{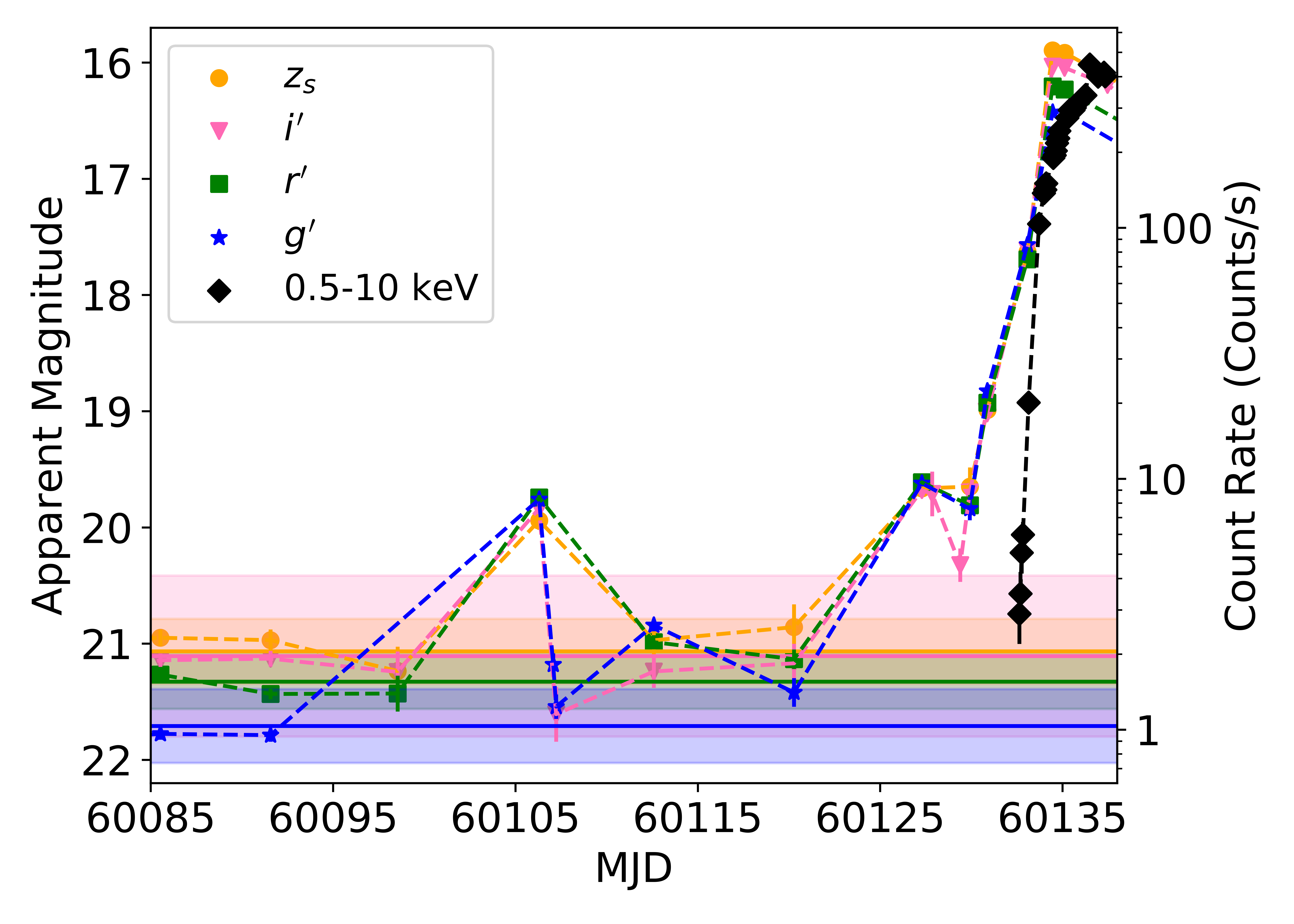}
    \caption{The rise of the 2023 outburst as observed with LCO and \nicer{}. The horizontal shades represent the $3\times~\text{RMS}$ extent from the mean quiescence levels calculated between MJD 60000 and MJD 60100, for the individual filters.}
    \label{fig:delay}
\end{figure}

\begin{figure*}
    \centering
    \includegraphics[scale=0.6]{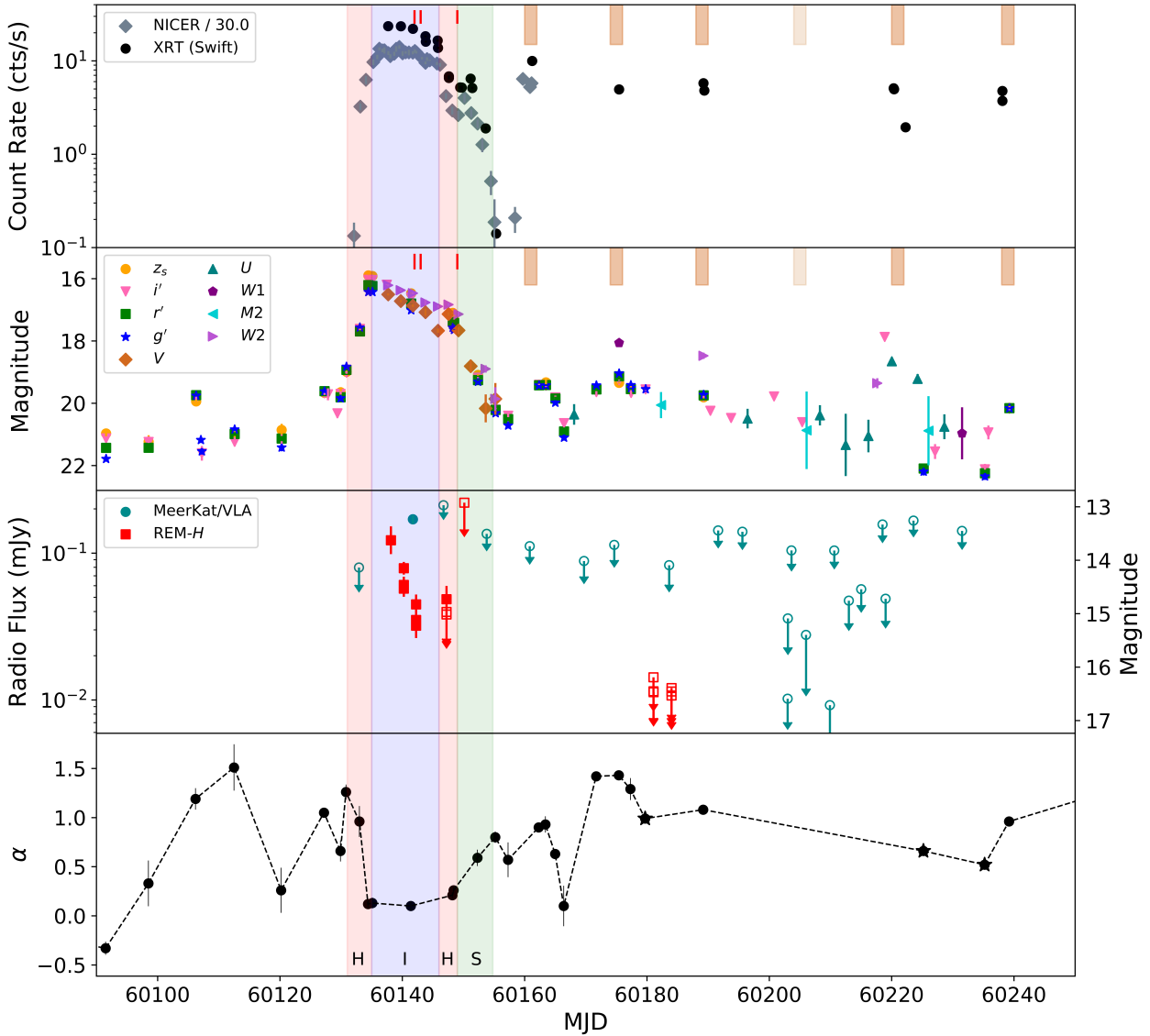}
    \caption{Top panel: X-ray lightcurves of \source{} with \nicer{} ($0.5-10$ keV) and \swift{}/XRT ($0.3-10$ keV). The \nicer{} lightcurve is reduced by a factor of 30 to plot it in the same frame as XRT. Second panel: Optical and UV lightcurves with LCO and \swift{}/UVOT, respectively. The light brown vertical bars dropping from the top in the first two panels represent the approximate epochs of the reflare peaks. The fourth epoch, marked $\sim 12$ days after the previous reflare, has a lighter shade as the peak of the reflare is not well constrained. The red vertical lines in these two panels represent the epochs when the optical polarization measurements were done with VLT. Third panel: Radio lightcurve made with MeerKAT/VLA data and IR ($H$-band) lightcurve with REM data. Solid markers represent detections whereas empty markers represent $3\times~\text{RMS}$ and $3~\sigma$ upper limits for radio and IR bands, respectively. Bottom panel: Evolution of the spectral index ($\alpha$) obtained by fitting the optical SED with a power law. The vertical patches that run across the four panels represent the boundaries of the spectral states as defined in \citet{rout25}. The `H', `I', and `S' are the initials for the hard, intermediate and soft states. The plateau phase discussed in this work broadly encompasses the intermediate state.}
    \label{fig:lc}
\end{figure*}

A brief brightening event was detected on MJD 60106.2, when the magnitudes decreased by more than 1 mag from the quiescence levels in the four LCO filters (Figure \ref{fig:delay}). This was about 26 days before the first significant X-ray detection with \nicer{} made on MJD 60132.1. Following this event, the fluxes decreased in all the filters to the quiescent levels (Figure \ref{fig:delay}). On MJD 60127.2, the source brightened up again and continued to rise, although after a small dip in the \ip{} band on MJD 60129.9, until the outburst peak on MJD 60134.4. The optical rise was detected 4.9 days before the first X-ray detection with \nicer{}. The estimation of the delay between the start of the X-ray rise and the start of the optical rise is quite uncertain, and depends on the sensitivity of \nicer{} and the gaps in the observations. If the actual X-ray rise started much before the detection (on MJD 60132.1), then the delay would be shorter. On the other hand, if the optical flux started to rise before the second brightening (MJD 60127.2), then the delay would be longer. Considering that the optical epoch previous to the rise was 7 days prior (MJD 60120.2), the actual delay between the start of the optical rise and the start of the X-ray rise could be anywhere between $4-12$ days. This assumes that the X-ray could start to rise from quiescence by at most one day before its detection, which may not be true.

After reaching the peak on MJD 60134.4, the optical flux gradually decayed until MJD 60150, forming a plateau (Figure \ref{fig:lc}). On MJD 60147.6 a flare was detected by UVOT in the \uvv{} and \uwt{} filters, about 2.6 days before it was detected in X-rays. This flare was missed by LCO because of gaps in the observations. After the flare, the optical flux declined rapidly until MJD 60157.3 marking the end of the main outburst. 

Before reaching the previous quiescence levels, the source started to show several high-amplitude rapid reflares over a period of more than two months. However, because of poor temporal sampling, it is difficult to trace the profiles of the reflares as well as count their exact numbers. Despite this, from the optical, UV, and X-ray lightcurves shown in Figure \ref{fig:lc}, the presence of at least six reflares can be ascertained. Approximately, they peak around MJDs 60163.4, 60175.4, 60189.2, 60200.8, 60218.9, and 60239.2. With the caveat that the peaks of the fourth and sixth reflares are not well constrained, the duration between each adjacent reflare is $\approx 12-14$ days, which is about twice that observed during the 2017 outburst \citep{jiminezibarra19}. The emission during the reflares follows a bluer-when-brighter pattern, with the X-rays being the highest amplitude, followed by UV and optical (see Figure \ref{fig:lc}).

\subsection{Spectral Energy Distribution} \label{sec:sed}

\begin{figure}
    \centering
    \includegraphics[scale=.5]{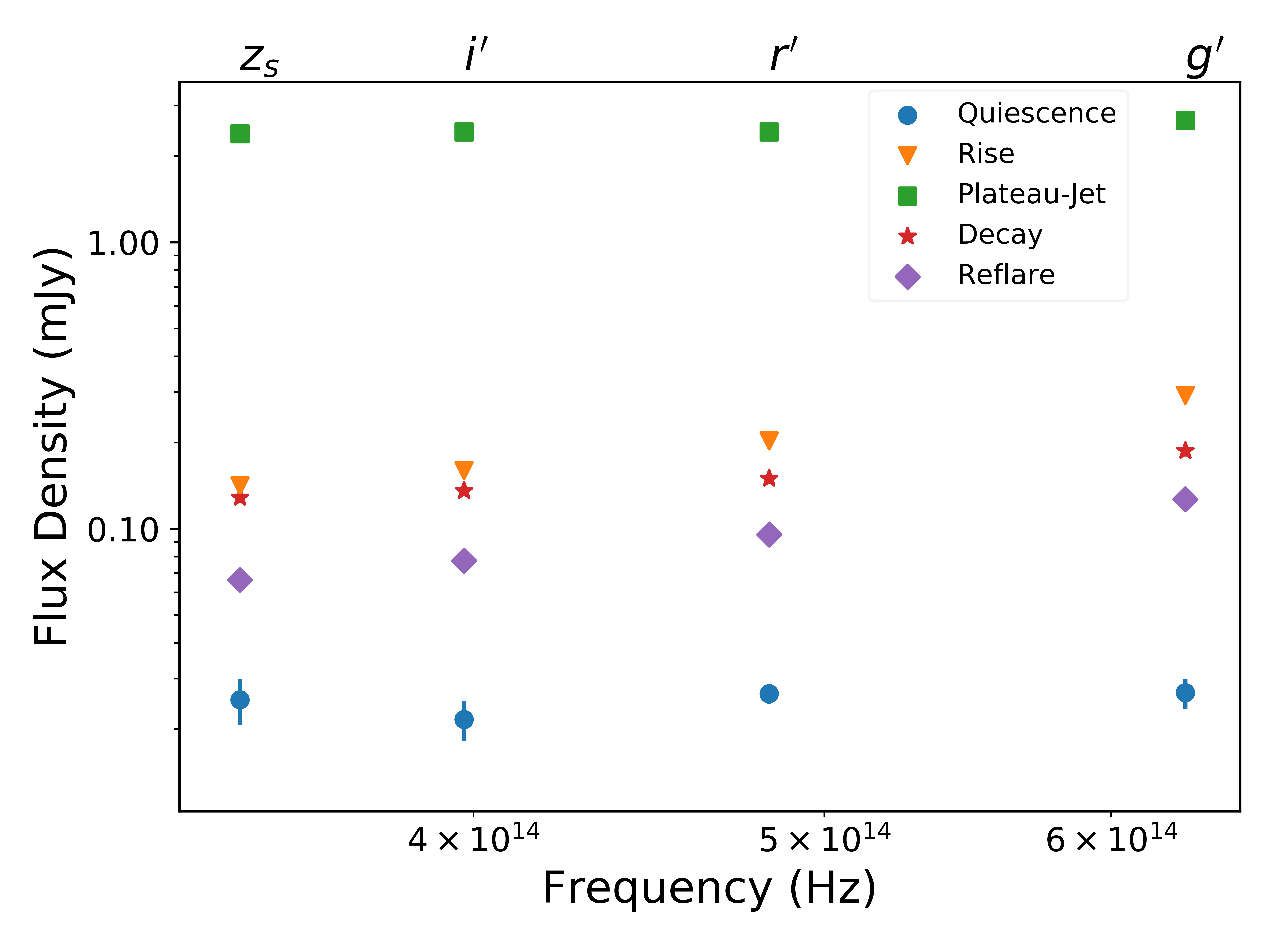}
    \caption{Five representative LCO SEDs during different stages of the outburst. The MJDs of the five epochs (from quiescence to reflare) are 60120, 60130, 60135, 60152, 60189.}
    \label{fig:optsed}
\end{figure}

\begin{figure*}
    \centering
    \includegraphics[scale=0.58]{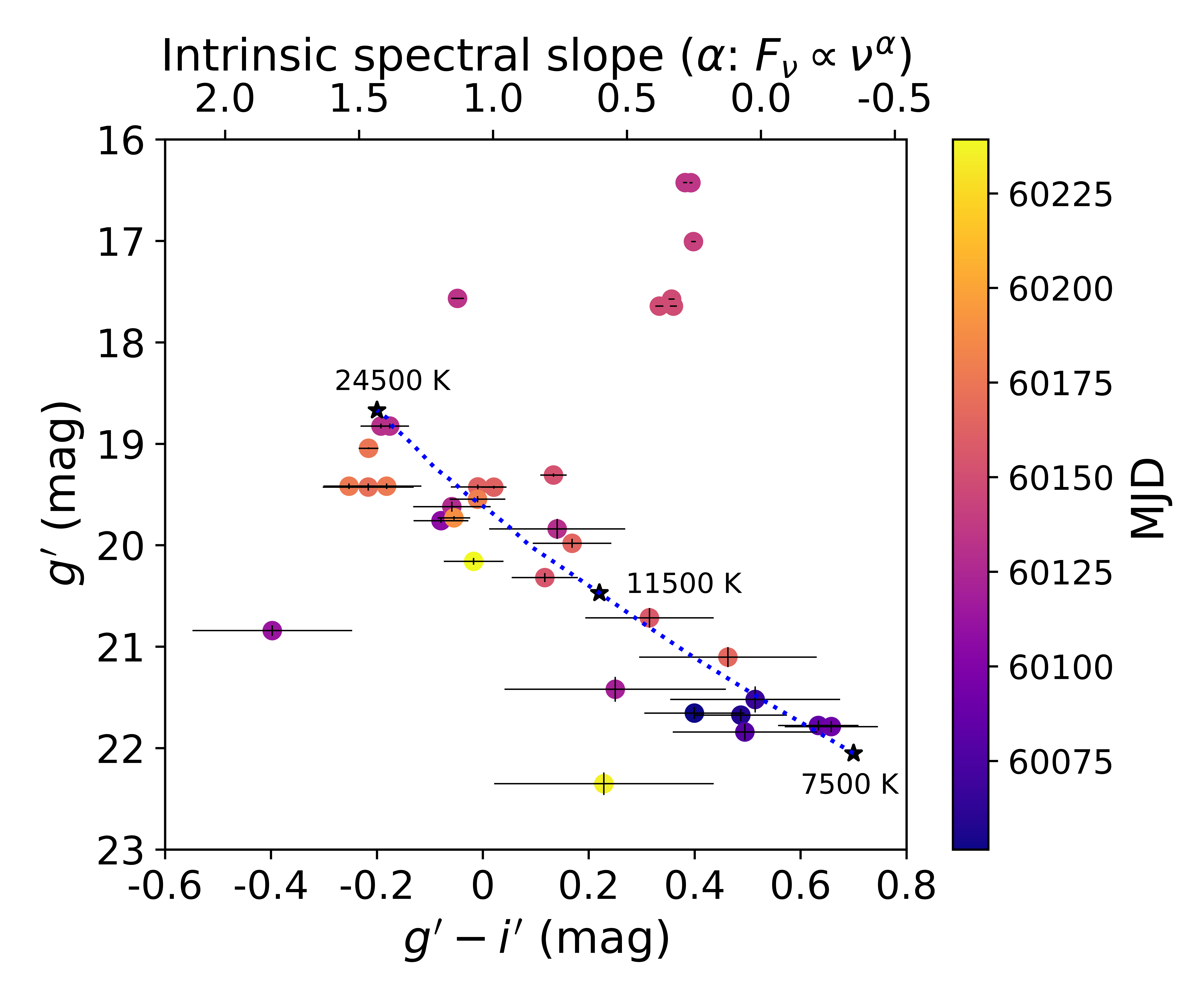}
    \includegraphics[scale=0.58]{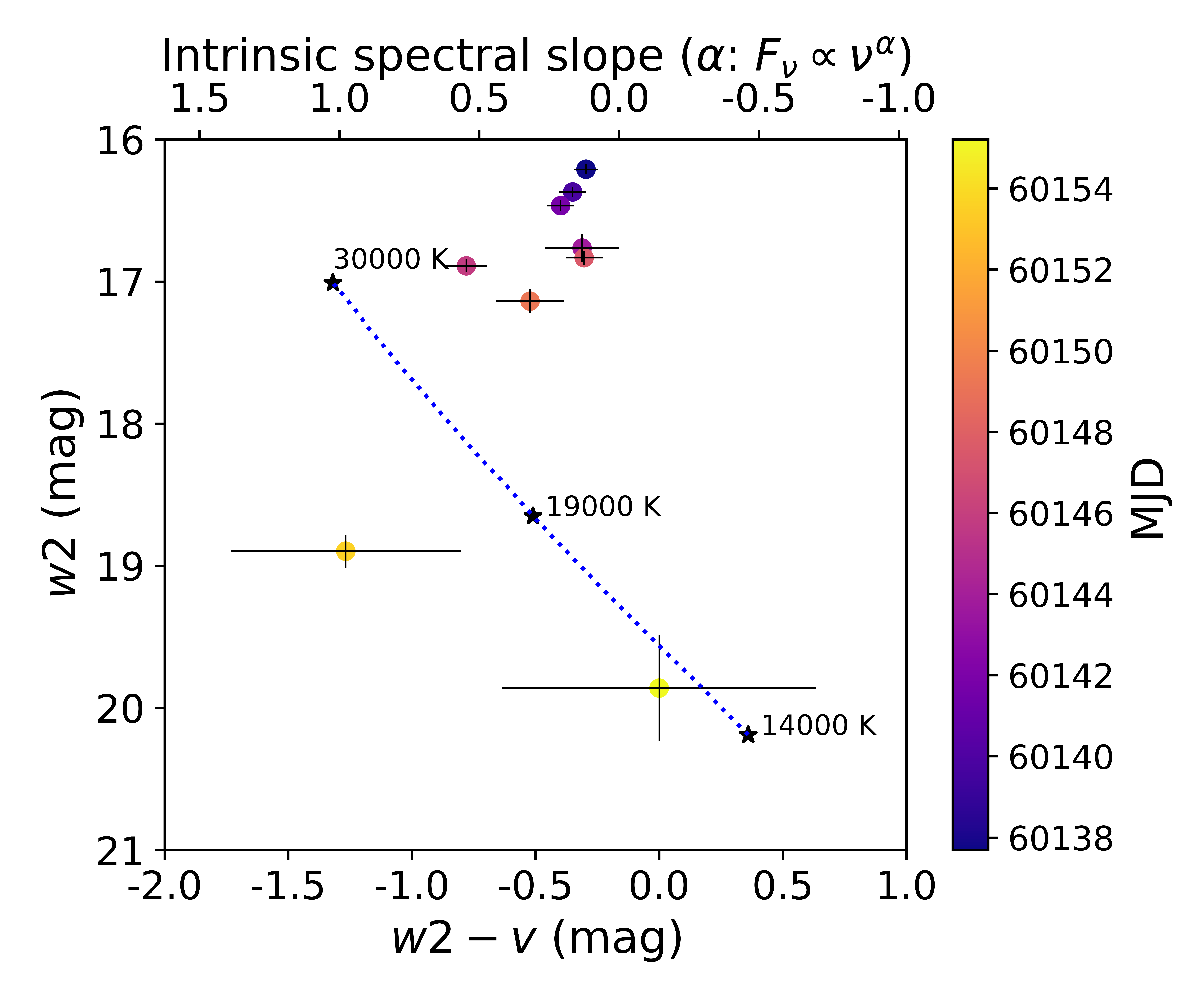}
    \caption{Color magnitude diagrams (CMD) with LCO \gp{} and \ip{} filters (Left), and UVOT \uwt{} and \uvv{} filters (Right). The blue dotted lines represent a blackbody model with different temperatures. The temperatures at three representative positions on the models are marked with stars. The top X-axis shows the de-reddened spectral slope.}
    \label{fig:cmd}
\end{figure*}

We studied the SEDs of \source{} to understand the optical emission mechanisms. To derive the SEDs from the apparent magnitudes, we first need the de-reddened fluxes. The optical extinction is linearly related to the equivalent hydrogen column density, \nh{} = $(2.62\pm0.02) \times 10^{21}$ cm$^{-2}$, obtained from X-ray measurements \citep{rout25}. The extinction is given by $A_V = \frac{N_H}{(2.87\pm0.12) \times 10^{21}} = 0.91\pm0.04$ mag \citep{foight16}. We used this value of $A_V$ along with the frequency-dependent absorption coefficients from the extinction law of \citet{cardelli89} to de-redden the calibrated optical magnitudes of the source. The final optical SEDs were constructed from observations taken in three or more filters within approximately one hour. These SEDs were fitted with a power law of the form $F_\nu \propto \nu^{\alpha}$, where $\alpha$ is the spectral index (Figure \ref{fig:optsed}). The evolution of $\alpha$ during the outburst, along with the multi-wavelength lightcurves, is shown in the bottom panel of Figure \ref{fig:lc}. During quiescence, the optical data exhibit considerable variability, both within individual filters and in the shape of the SEDs. In some cases, the SEDs do not conform to a power law (see Figure \ref{fig:optsed}). The SED may trace the curved peak of the Planck function or be contaminated by emission from the secondary star, making the power law a poor approximation. Nevertheless, it is evident that the overall slope of the SED during quiescence varies over a wide range, between $-0.33 \pm 0.07$ and $0.33 \pm 0.23$ (1$\sigma$). During the optical flare on MJD 60106.2, $\alpha$ increases to $1.19 \pm 0.11$ and remains there (within errors), even as the flux decreases in the next epoch on MJD 60112.5. On MJD 60120.2, the fluxes in all filters return to near-quiescent levels, and the SED becomes flat, with $\alpha = 0.26 \pm 0.23$. MJD 60127.2 marks the rise of the main outburst, characterized by a steepening of the SED ($0.66 < \alpha < 1.26$) until MJD 60134.4, near the peak, when the index suddenly drops to $0.12 \pm 0.04$. Afterward, $\alpha$ gradually increases in anti-correlation with the flux, reaching $0.57 \pm 0.18$ on MJD 60157.3, when the main outburst ends. In the first reflare that follows the main outburst, the spectral index tracks the flux, rising to $0.93 \pm 0.08$ and then dropping to $0.10 \pm 0.21$. In subsequent reflares, however, the sampling is sparse, making it difficult to trace the evolution of the SED. Generally, around the peak of the reflares, the SED remains steep with $\alpha \sim 1.0$ (Figure \ref{fig:optsed}). Observations on MJDs 60225.2 and 60235.2 may be exceptions, where $\alpha \sim 0.6$ when the optical fluxes are at quiescent levels. However, these SEDs are based on only two filters (the points marked with black stars in Figure \ref{fig:lc} have only two filters and thus no error bars), making it difficult to draw firm conclusions.

\subsection{Color Evolution} \label{sec:cmd}

The color-magnitude diagram \citep[CMD;][]{maitra08}, similar to the hardness-intensity diagram in X-rays, is a useful tool for ascertaining state changes and delineating the different emission mechanisms in X-ray transients \citep{russelld11}. In Figure \ref{fig:cmd}, we show the CMD with LCO bands (\gp{} versus \gp{}$-$\ip{}) in the left panel and with UVOT bands (\uwt{} versus \uwt{}$-$\uvv{}) in the right panel. \swift{} observations were triggered after the detection of the optical rise, so we have data from the outburst peak until the end of the main outburst. Before the optical activity began, during the quiescent stages, \source{} occupied the bottom right position with redder colors ($\alpha \lesssim 0$). With the optical rise, the source became bluer ($\alpha \gtrsim 0$) until MJD 60133.0, just before the outburst peak, when the color becomes slightly redder. In the following six observations, until MJD 60152.3, the colors remained significantly red. These observations stand out as a horn in the upper right part of each CMD (Figure \ref{fig:cmd}). In the rest of the observations, the color evolution follows the normal diagonal track in the CMD.   

We modeled the CMDs with a uniform-temperature blackbody of constant area as it cools and heats \citep{maitra08, russelld11}. The flux normalization of the model depends on several parameters such as the masses of the two stars, the size and inclination of the disk, the distance to the source, and other factors, most of which are degenerate and poorly constrained for \source{}. Hence, the flux normalization is also poorly constrained, and we opted to use a multiplicative factor to overlap the model with the data. This factor is proportional to the projected area of the blackbody, which depends on the aforementioned uncertainties and which we assumed to remain constant. For UVOT filters, we used the same parameters, albeit with a different range of temperatures that is consistent with UV emission. It is interesting to note that the disc is already at a high temperature of $\approx 7500$ K before the start of the outburst. Before deviating towards redder colors from the blackbody track near the peak, the temperature increases to $\approx 24500$ K in the optical bands and to $\approx 30000$ K in the UV.

\subsection{X-ray - Optical/UV correlations} \label{sec:xuoircorr}

\begin{figure*}
    \centering
    \includegraphics[scale=0.55]{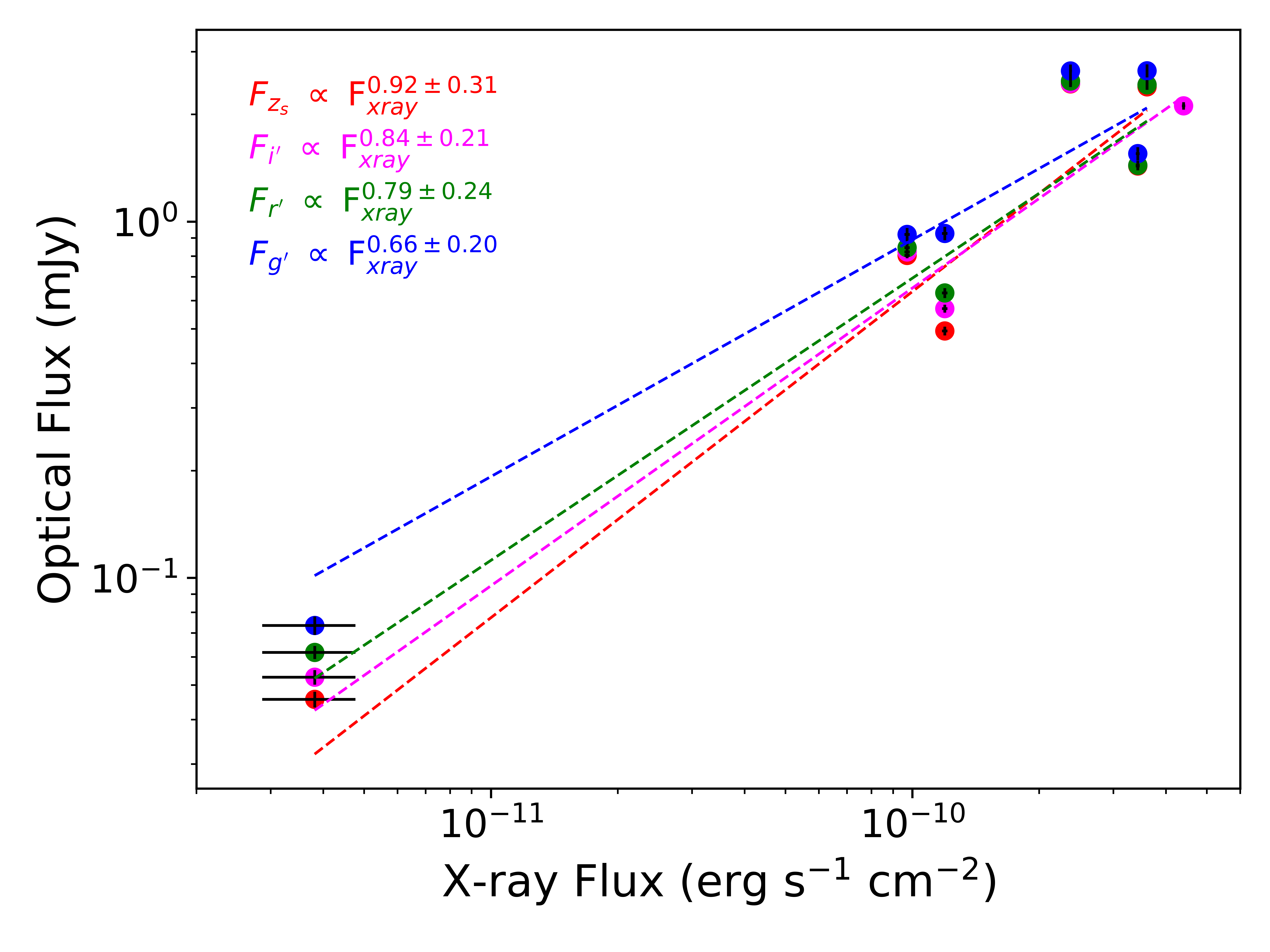}
    \includegraphics[scale=0.55]{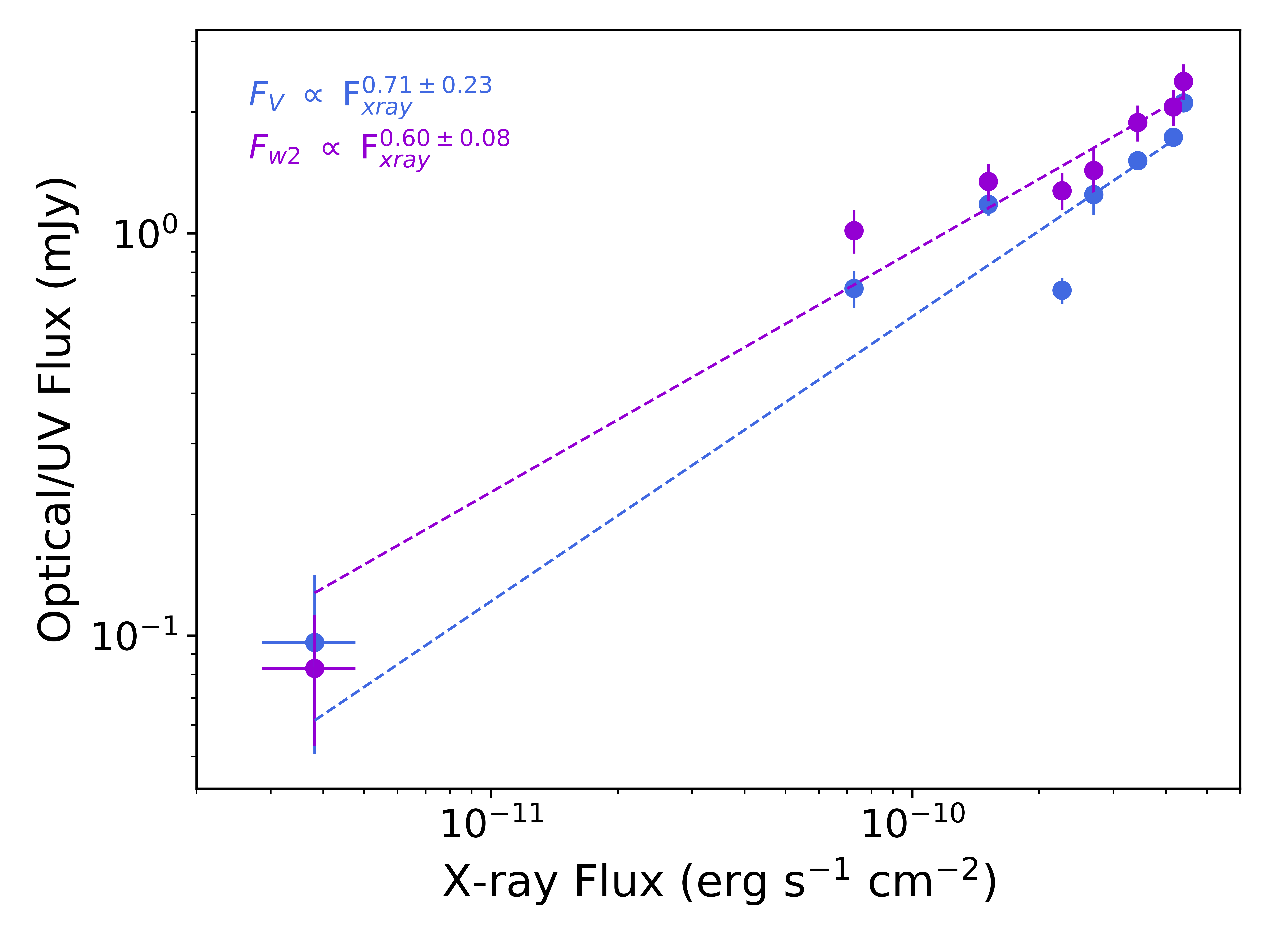}
    \caption{Left: X-ray/Optical correlation with the four filters of LCO. Right: X-ray/Optical-UV correlation with the two filters of UVOT. The unabsorbed X-ray flux is in the $2-10$ keV range from fits to the \nicer{} spectra. The dashed lines represent the best fitting powerlaw models.}
    \label{fig:xoptcorr}
\end{figure*}

We studied the correlation between the optical/UV and quasi-simultaneous X-ray fluxes to determine the dominant optical and UV emission mechanism \citep{russelld07}. Of all available observations, data from six LCO epochs were found to be within 24 hours of an X-ray observation with \nicer{} in the hard or intermediate states. Similarly, eight observations were found to be quasi-simultaneous between \nicer{} and \swift{}/UVOT (only \uvv{} and \uwt{} filters) during the hard or intermediate states. Only one and two observations with LCO and UVOT, respectively, were found to be quasi-simultaneous with \nicer{} during the soft state. Therefore, we did not carry out a correlation study in the soft state. The unabsorbed X-ray fluxes in the $2-10$ keV range were calculated from spectral fits to the \nicer{} data \citep{rout25}. 
The fluxes in both optical/UV and X-ray bands span about two orders of magnitude, and there is a clear positive correlation, albeit with significant scatter. We fitted the correlations with a power law using the ODR algorithm\footnote{\url{https://docs.scipy.org/doc/scipy/reference/odr.html}}. Figure \ref{fig:xoptcorr} shows the correlations for six optical/UV filters (LCO - \zs{}, \ip{}, \rp{}, \gp{} and UVOT - \uvv{}, \uwt{}) along with the best-fitting models (dashed lines). The power law index, $\beta$ ($F_{\text{opt/uv}} \propto F_{\text{xray}}^{\beta}$), monotonically increases with increasing wavelength of the optical/UV filters, though with large uncertainty. While it is $0.60 \pm 0.08$ for the \uwt{} filter, it increases to $0.92 \pm 0.31$ for the \zs{} filter. There is systematically more scatter in the longer wavelength bands, which is evident from the corresponding levels of uncertainty (Figure \ref{fig:xoptcorr}). 

We investigated the significance of the correlation between the wavelengths and $\beta$ using a similar method as used in Section \ref{sec:ana-delay}. First, we fitted the correlation by a constant model and then by a line, with a non-zero slope and intercept, using $\chi^2$ minimization. The $\Delta\chi^2$ between the two models is 2.22 for one additional degree of freedom, indicating that they are not significant at the $5\%$ or $10\%$ level. This suggests that while there may be a correlation, it is not highly significant.

\subsection{Optical Polarization} 

\begin{table*}[]
\caption{Results of the VLT/FORS2 ($BVRI$ filters) polarimetric campaign. All of the polarization levels and angles are corrected for instrumental polarization. The interstellar polarization has also been subtracted, by means of a group of reference stars in the field. Upper limits are indicated at a $99.97\%$ credible interval, and the rest of the uncertainties are $\pm1 \sigma$ credible intervals.}            
\label{tab:pol}      
\centering                       
\begin{tabular}{c |c |c |c |c |c |c| c |c |c}       
\hline               
 \multirow{2}{*}{Epochs} & \multirow{2}{*}{MJD} & \multicolumn{2}{c|}{$B$} & \multicolumn{2}{c|}{\uvv{}} & \multicolumn{2}{c|}{$R$} & \multicolumn{2}{c}{$I$} \\   

 & & $P$ (\%) & $\theta$ ($^{\circ}$)& $P$ (\%) & $\theta$ ($^{\circ}$) & $P$ (\%) & $\theta$ ($^{\circ}$) & $P$ ( \%) & $\theta$ ($^{\circ}$)\\
% \hline

\hline                       
1 & 60142.1 &$0.61^{+0.07}_{-0.09} $ &$80.49^{+3.54}_{-3.44} $  & $0.55^{+0.05}_{-0.06} $ & $85.63^{+2.90}_{-2.83}$ &$0.57^{+0.04}_{-0.05} $   & $82.17^{+2.27}_{-2.26}$  &  $0.60^{+0.04}_{-0.05}  $ & $63.36^{+2.24}_{-2.10} $   \\
%\hline  

\hline
2 & 60143.1 & $ 0.29^{+0.07}_{-0.08}$ & $99.09^{+7.09}_{-7.13} $ &  $0.34^{+0.05}_{-0.06}  $ &$83.24^{+4.47}_{-4.45} $ & $0.51 ^{+0.04}_{-0.05} $ & $75.88^{+2.51}_{-2.45} $ & $<0.17 $ & $87.05^{+36.58}_{-34.54} $ \\
%\hline  

\hline

3 & 60149.0 & $<0.56 $ & $137.96^{+11.48}_{-13.19}$ & $<0.4 $ & $144.42^{+16.07}_{-29.68}$ & $1.03\pm 0.06 $ & $176.83^{+1.77}_{-1.79}$ & $0.63 \pm 0.06 $& $192.20\pm 2.75$ \\
%\hline  

\hline

\end{tabular}
\end{table*}

\begin{figure}
    \centering
    \includegraphics[scale=0.3]{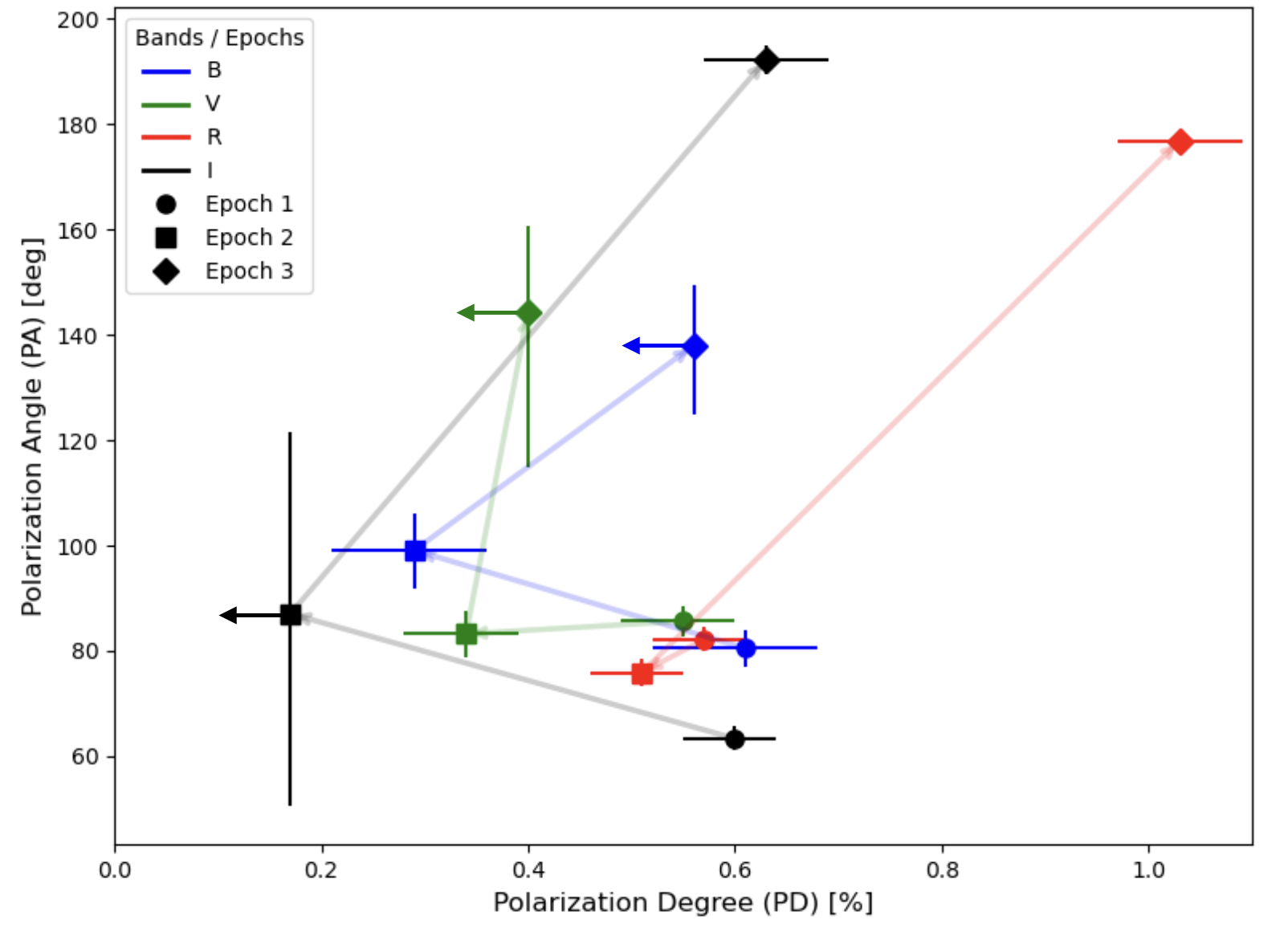}
    \caption{Optical polarization degree ($P$) versus polarization angle ($\theta$) for the three epochs. A clear rotation of $\theta$ by approximately $\sim100^\circ$ is observed in the $R$ and $I$ bands between the first two epochs and the third. Upper limits on $P$ are reported at the $99.97\%$ credible interval, while all other uncertainties are quoted at the $\pm~1\sigma$ credible level.}
    \label{fig:polevo}
\end{figure}

The results of our analysis of the optical polarization observations performed with FORS2 are given in Table \ref{tab:pol}. The level of linear polarization, after the correction for the reference stars, is always $\leq 1\%$ in all bands and at all epochs. In the first epoch, $P$ is constant at a level of $\sim 0.6\%$ at all frequencies, and the polarization angle, $\theta$, is approximately constant at $\sim 80^{\circ}$ (with a slight but significant deviation in the $I$ band). In the second epoch, $P$ reduces at all frequencies, while $\theta$ maintains similar values. A significant variation in polarization is observed in the third epoch. The level of polarization becomes undetectable in the $B$ and \uvv{} bands, with constraining upper limits. In contrast, the $P$ values in the $R$ and $I$ bands are significantly higher than those in the second epoch, with a swing in $\theta$ of $\sim 100^{\circ}$\footnote{This is if the electric field vector rotates in the anti-clockwise direction. If the vector rotates instead in the clockwise direction, the swing in polarization will be $\sim75^\circ$.} in both bands. This evolution of polarization properties in the three epochs can be visualized in Figure \ref{fig:polevo}.

\subsection{Radio Detection}

\begin{figure}
    \centering
    \includegraphics[width=1\linewidth]{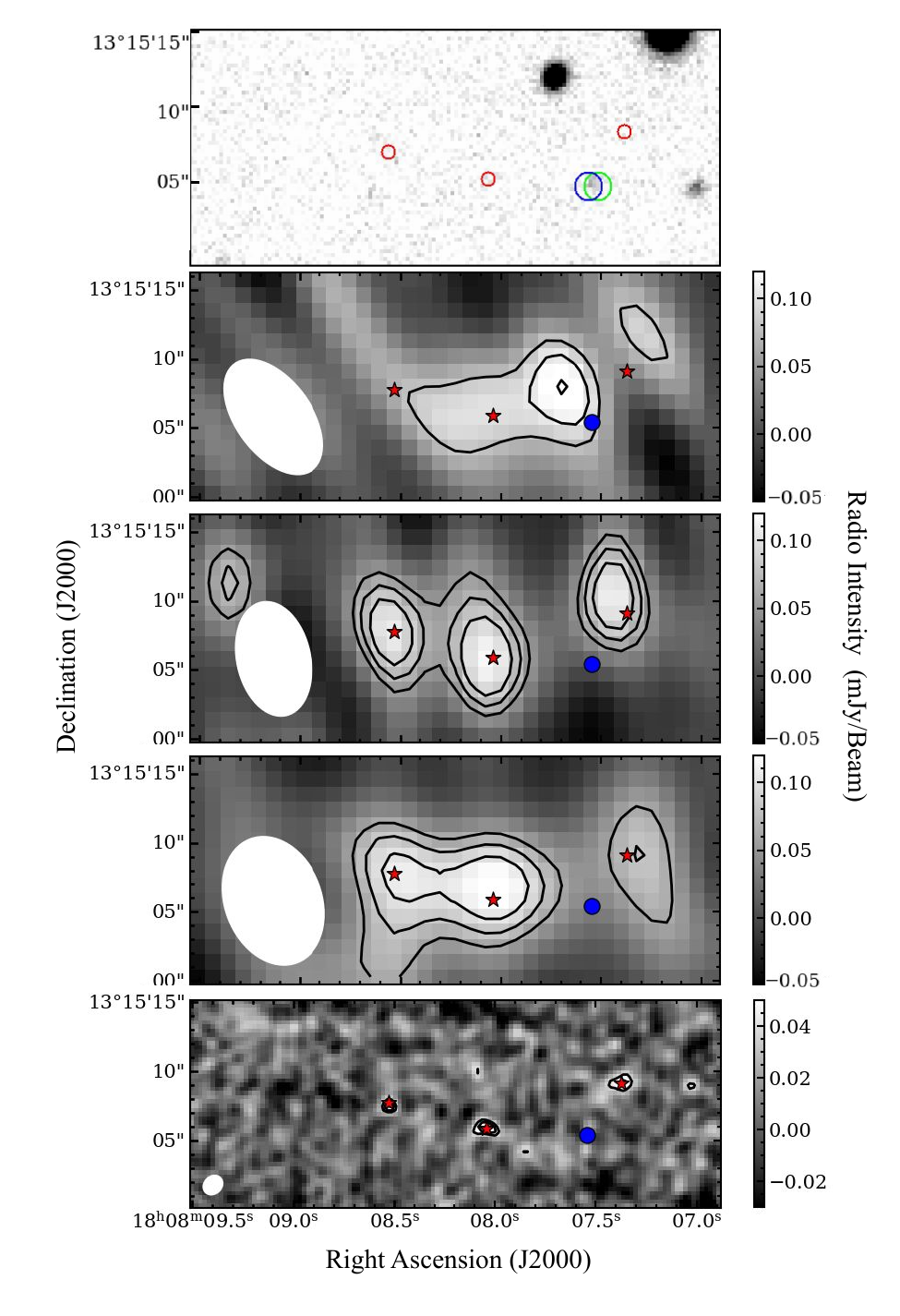}
    \caption{Multi-frequency images of \source{} and the surrounding region. The top panel shows an optical (i$^\prime$-band) image taken with the LCO on 2024 August 2 (MJD 60524); the second panel shows a MeerKAT ($\sim$\,1.3\,GHz) image taken on 2023 July 16 (MJD 60141); the third panel shows a MeerKAT ($\sim$\,1.3\,GHz) image taken on 2023 July 28 (MJD 60153); the fourth panel shows a MeerKAT ($\sim$\,1.3\,GHz) image taken on 2023 September 4 (MJD 60191); the bottom panel shows a VLA ($\sim$\,1.5\,GHz) image taken on 2023 October 2 (MJD 60219). The contours in the four radio images represent $3\sigma$, $4\sigma$, and $5\sigma$ levels. The open red circles and solid red stars show the radio positions of the field sources in the optical and radio images, respectively. Similarly, the open blue circle and the solid blue circle show the optical position of the source in the optical and radio images, respectively. The open green circle in the optical image corresponds to the UVOT position of the source. The noise levels for the four radio images, from top to bottom, are 27 $\mu$Jy/beam, 18 $\mu$Jy/beam, 20 $\mu$Jy/beam, and 13 $\mu$Jy/beam. The three field sources are consistent with having constant fluxes in the four epochs, to $<2\sigma$ of the mean level. The first MeerKAT image shown in the second panel corresponds to the only radio detection of MAXI J1807$+132$. }
    \label{fig:radio}
\end{figure}

Only one MeerKAT observation (2023 July 16, MJD 60141) showed significant evidence for radio emission originating from MAXI J1807$+$132 (shown in the second panel of Figure \ref{fig:radio}). For this detection, we extracted the source flux (with \textsc{imfit}) using a synthesized beam-shaped Gaussian. We measured the root mean square (RMS) noise from a nearby source-free region using a circular extraction aperture with an area equal to $\sim$\,100 synthesized beams. We observed an unusually large offset between the radio and optical positions (${\sim}\,3^{\prime\prime}$). Comparably large astrometric errors have been seen in past comparisons of radio-infrared positions\footnote{See obit development memo series no.~6: \url{https://www.cv.nrao.edu/~bcotton/ObitDoc/SelfCalAstro.pdf}}, where group delay errors were found to be the most likely culprit; although not certain, it is possible that these errors also affect our radio observations. Regardless, the temporal coincidence of the detection with the peak of the outburst and the fact that the new component is at a different position from the three persistent sources are consistent with the source emission originating from \source{}. 

For the remaining non-detections, we performed forced aperture photometry by extracting the peak flux density using a synthesized beam-shaped Gaussian fixed at the optical position. Then we added the forced aperture flux densities and $3\times$ the RMS noise value, and adopted these as $3\sigma$ flux upper limits. Therefore, our upper limits consider both noise-driven uncertainty and confusion from the nearby field sources. The radio fluxes of MAXI J1807$+$132 are shown in Figure \ref{fig:lc}.

\section{Discussion} \label{sec:discussion}

We conducted a multi-wavelength analysis of the NS LMXB \source{} during its latest outburst in 2023, using data from several ground and space-based observatories. Thanks to the prompt initiation of X-ray observations following the optical rise, we were able to detect the X-ray rise early in the outburst. Despite some uncertainties, we detected a clear delay of $\sim 4-12$ days between the X-ray and optical rises. The outburst was characterized by a rapid rise, a gradual decay (the plateau phase), and, after a brief flare, a sharp drop to near-quiescence levels. We also detected signatures of jet synchrotron emission in the optical and infrared (OIR) bands throughout the plateau phase. Following the outburst, a series of high-amplitude reflares ($\sim 2$ orders of magnitude in flux), with a quasi-periodicity of $\gtrsim 12$ days, occurred over a period of three months.

\subsection{Optical precursor and X-ray delay}

\source{} undergoes a slow and long-term rise during the quiescence that began $\sim400$ days before the main outburst (Figure \ref{fig:slowrise}). This long-term trend is expected from the DIM as matter accumulates gradually in the accretion disk, leading to an increase in surface density and temperature \citep[e.g.,][]{dubus01}. The rise rates for the bluer bands ($\sim0.3$ mag/year) are steeper than the redder bands ($\sim0.1$ mag/year) consistent with previous observations of LMXBs \citep[e.g.,][]{bernardini16a, koljonen16}.  A precursor to the main outburst was detected on MJD 60106.2, i.e., $\sim 26$ days before the first X-ray detection. This enhanced optical flux, however, decayed down to near-quiescent levels for a couple of weeks. The commencement of the fast optical rise, when the heating wave sweeps through the disc, likely began somewhere between MJD 60120.2 and 60127.2, i.e., about $4-12$ days before the X-ray rise. The precursor delay ($\sim 26$ d) is too long to be associated with viscous timescales of the disc, which are of the order of a few days. It probably appears as a result of some other activity that leads to the enhanced accretion in the next 15 days. Such a precursor was also detected in the accreting millisecond pulsar SAX J1808.4--3658 12 days before an optical rise \citep{goodwin20}. This suggests that the precursor could be a generic feature of LMXB outbursts, although their origin is unclear. It is unlikely to be similar to the misfired outburst in Cen X-4 \citep{baglio22}, where the temperature does not reach the hydrogen ionization level. In both \source{} (Figure \ref{fig:cmd}) and SAX J1808.4--3658 \citep{baglio20} the disc temperature exceeded $\sim 7000$ K.  One possibility could be an enhanced mass transfer due to some stellar activity in the companion, which results in an optical flare. Alternatively, some instabilities can arise due to the impact of the accretion stream, or from spiral waves in a dense outer disk, resulting in optical activity.     

One key observable of the DIM is the delay between the rise in X-ray and optical flux at the beginning of an outburst. According to the DIM, the outburst begins when, due to increasing viscosity, the disk temperature rises to a level where hydrogen starts to ionize, i.e. $\gtrsim$ 6500 K. This is a temperature at which a blackbody peaks in the blue region of the optical spectrum. As the heating front reaches the inner disk, the temperature increases to levels where X-rays are emitted. However, even before the direct disk emission, the X-rays start to emanate because of Comptonization of the optical/UV emission in the hot flow. Therefore, the rise of X-ray emission is delayed with respect to the rise of optical (blue) emission at the beginning of the outburst. The exact duration of the delay, however, depends on a number of factors related to the time it takes (i) the heating wave to travel from the ignition radius to the inner disc, (ii) the inner disc to propagate towards the neutron star, and (iii) the mass accretion rate to rise in the hot flow. If the ignition occurs at a smaller radius, then the heating front reaches the inner disk faster than in the case where the ignition occurs at a larger radius. These two cases are termed inside-out and outside-in outbursts, respectively \citep{dubus01}. However, this does not imply that inside-out outbursts have a smaller delay \citep{hameury20}.  

Determining the type of outburst, i.e., inside-out or outside-in, requires constraining the ignition radius, which depends on the size of the disk and the mass transfer rate from the companion, both of which are quite difficult to constrain \citep{smak84}. Moreover, the $4-12$ day delay between the X-ray and optical rise measured in \source{} is much greater than the time it takes for the heating to reach the inner edge of the disk, no matter where it originates in the disk. For instance, the heating front could take less than a day to reach the inner edge if it is an inside-out outburst and/or if it is a short-period system. The delay is also a function of the viscosity parameter \citep{hameury97}. Delays of the order of a few to several days can be explained if the disk is truncated at the beginning of the outburst, either due to magnetic pressure \citep{livio92} or evaporation \citep{meyer94}. The heating front would then have to stop at the truncated inner edge of the disk, which would then move inward at a viscous time scale, which is longer than the propagation time of the heating front \citep{menou99, dubus01}. From the X-ray study of the source, it was indeed found that the disc was highly truncated at the beginning \citep{rout25}. Viscous time-scales for the propagation of matter or turbulence from the optical to X-ray emitting region have been attributed to a few BH LMXBs, namely GRO J1655--40 \citep[6 d;][]{hameury97}, LMC X-3 \citep[$5-10$ d;][]{brocksopp01}, GX 339--4 \citep[$15-20$ d;][]{homan05a}, XTE J1118$+480$ \citep[$>4$ d;][]{zurita06}, 4U 1957$+$11 \citep[$2-14$ d;][]{russelld10a}, Swift J1910.2--0546 \citep[6 d;][]{degenaar14, saikia23}, etc. However, it is important to note that the viscous timescales measured in these studies were done during the outburst, when the whole disc is ionized and the timescale is dominated only by the viscous timescale. This is not directly comparable with the delays of the X-ray rise at the start of the outburst, which depend on many other factors mentioned above. 

\subsection{UV-Optical-IR (UVOIR) Emission processes} 

The outburst of \source{} shows a very fast rise, so much so that it reaches the peak about two days after its first detection in X-rays. Figure \ref{fig:lc} shows the boundaries of the spectral states as defined in \citet{rout25} on the multi-waveband lightcurves with different colors. The source transitions from the hard state to the intermediate state between the peak and the end of the gradually declining plateau phase on MJD 60146. Following this, the state becomes hard for a few days marked by a sudden drop in flux and an increase in the X-ray variability and hardness. On MJD 60150, the source transitions to the soft state with a short flare in the X-rays. The optical/UV lightcurves also follow the same trend as the X-ray lightcurve, although the decay in the intermediate state is slightly shallower than the X-ray decay. It is during the intermediate and hard states between MJD 60134 to 60150 that we detected signatures of jet synchrotron emission at optical wavebands. The OIR emission may have contributions from the accretion disk (viscous or irradiated), secondary star, hot spots, hot flow, or jet,  making the identification of the dominant mechanisms difficult \citep{shakura73, cunningham76, vanparadijs94,corbel02,russelld06,veledina13}. Therefore, we employed a number of diagnostic tools to identify the emission mechanism which are discussed below.

\subsubsection{Color evolution - SED and CMD} \label{sec:sedcmddis}

We fitted optical SEDs with a power law ($F_{\nu} \propto \nu^{\alpha}$) parameterized by the spectral index, $\alpha$, along with a normalization. The evolution of the spectral index is shown in Figure \ref{fig:lc}. Although for most of the outburst, the SEDs are bluer (i.e., steeper with $\alpha \approx 1.0$) when brighter, they become flat ($\alpha \approx 0$) during the plateau phase. The steep spectral indices during the rise and reflare peaks are consistent with the disk emission, both viscous and irradiated \citep{hynes05}. The shallow index range ($\sim 0$-0.4) at the peak and initial decay (plateau), on the other hand, is consistent with a multi-temperature blackbody from the viscous accretion disk (where $\alpha \sim 1/3$ is expected) or a self-absorbed jet \citep[which can be flat or slightly inverted; $\alpha \sim 0$-0.5 e.g.,][]{fender01a, russelld13b, russellt14}. If the flattening in the SED is caused by the disk emission, then the temperature must be lower than that measured. This becomes apparent from the CMD modeling, as discussed in the following. 

The reddening of the spectral index detected in the SEDs also becomes apparent in the CMD. The \gp{}$-$\ip{} color starts to deviate from the diagonal track and becomes redder just before reaching the peak of the outburst (Figure \ref{fig:cmd}). Although UVOT missed the rise of the outburst, it started observing from the peak until the end of the main outburst. Quite remarkably, the UVOT CMD also shows that the \uwt{}$-$\uvv{} color is considerably red near the peak and moves toward bluer colors as the outburst decays (Figure \ref{fig:cmd}). The diagonal track can be modeled by a blackbody function, and it is represented by a blue dotted line. The points along this model are dominated by the irradiated-disk emission, whereas the outlier points have a significant contribution from some other source \citep[e.g.,][]{russelld11}. This is based on the principle that for an irradiated disk scenario the optical emission is dominated by different parts of the Planck's blackbody function at different temperatures. At low temperatures, the optical emission comes from the peak or the Wien's tail, hence a flat or red SED (high \gp{}$-$\ip{}), while at higher temperatures the emission is dominated by the Rayleigh-Jeans tail resulting in a bluer SED (low \gp{}$-$\ip{}). Whenever there is a significant contribution from any other source, the observed colors tend to deviate from the model. It has often been seen that a significant contribution from jet synchrotron emission results in a red-ward shift of the colors away from the blackbody track \citep[e.g.,][]{baglio20}. The six points in \source{} starting at the peak of the outburst (top right corner in Figure \ref{fig:cmd} between MJD 60134.4 to MJD 60152.3) represent such an excursion possibly arising from the jet.   

\subsubsection{X-ray vs. Optical/UV correlation} \label{sec:corrdis}

If the UVOIR emission is dominated by the viscous disk, then the correlation index $\beta \sim 0.3$ ($F_{\text{opt}} \propto F_{\text{xray}}^{\beta}$). For an irradiated disk, this would be $\beta \sim 0.5$ \citep{vanparadijs94, frank02}. \citet{russelld06} found a global OIR/X-ray correlation index of $\sim 0.6$ for both NS X-ray binaries (NSXBs) and BH X-ray binaries (BHXBs) in the hard state. This correlation can be adequately explained by an X-ray reprocessing model for BHXBs, and also for NSXBs at low luminosities. At luminosities $\gtrsim 10^{36}$ erg s$^{-1}$, the jet may dominate the OIR flux for atoll sources \citep{russelld07}. There have not been many studies dedicated to finding a correlation when the OIR flux is dominated by jet emission. This correlation would be similar to the radio/X-ray correlation if the optically thick part of the synchrotron spectrum is flat and the spectral break occurs at higher frequencies, i.e. in the optical regime \citep[e.g.,][]{diaztrigo18}. Several studies have been dedicated to finding a radio/X-ray correlation, and a range of values have been estimated over the years. \citet{migliari06a} found a very steep radio/X-ray correlation of 1.4. \citet{gallo12}, on the other hand, reported a bimodality in slopes corresponding to radio-loud ($\beta \sim 0.6$) and radio-quiet ($\beta \sim 1$) jets. \citet{gallo18} even reported a shallower slope of 0.71 for atolls, and even a lower value of 0.44 for all NSXBs. Probably limited by detection sensitivities, there seems to be no global radio/X-ray correlation in NSXBs \citep[e.g.,][]{tudor17,vandeneijnden21}. Moreover, NSs are inherently more complex than BHs owing to a surface that spins and anchors magnetic fields which can result in more scatter in correlations. 

The studies done so far suggest that the correlation between UVOIR and X-ray flux becomes steeper (i.e. increases from $\sim 0.3$ to 1.4) as one moves from a viscous disk to irradiated disk and then to jet emission. For instance, the mid-IR flux, dominated by jet emission, was recently found to correlate with the X-ray flux with a steep $\beta \sim$ 0.82 \citep{john24}. The correlation indices for \source{} in the \uwt{}, \gp{}, and \uvv{} bands ($\beta \sim 0.6$) are consistent with X-ray reprocessing. The longer wavelength bands (\rp{}, \ip{}, and \zs{}), on the other hand, show a somewhat steeper correlation with X-rays ($\beta \gtrsim 0.6$). This suggests a possible contribution of the jet in the \rp{}, \ip{}, and \zs{} bands. As shown in Section \ref{sec:xuoircorr}, although not highly significant, there is a hint of a positive correlation between $\beta$ and wavelength that is opposite to that expected from a viscous heated disk \citep[e.g.,][]{armaspadilla13,patruno16}.

\subsubsection{Infrared and radio detections}

During the hard and intermediate states of X-ray binaries (XRBs), outflows in the form of jets are usually detected, and these jets are then quenched by about 3 orders of magnitude as the source transitions to a soft state \citep[e.g.,][]{russellt14, carotenuto21}. Jet emission is clearly identified at radio wavelengths, while its contribution in the millimeter and IR regimes is also well known \citep[e.g.,][]{rout21b}. Our radio observation campaign with MeerKAT and VLA results in a detection at the source position (Figure \ref{fig:radio}) during the intermediate state on MJD 60141 (2023 July 16). This is the only significant detection of the 22 observations. Moreover, observations in the $H$ band carried out with REM, reveal strongly variable emission during the intermediate and hard states (Figure \ref{fig:lc}). The radio detection and variability of the IR emission during the plateau phase of the outburst likely results from a brightened jet, which would, in turn, also explain the reddening of the optical color. 

As the radio flux at $\sim 0.17$ mJy is about a factor of 10 below the IR and optical flux, the optically thick component of the synchrotron emission is inverted. Due to the lack of data in the mid-IR wavebands, it is not possible to constrain the break frequency. However, assuming the \zs{} band to represent the upper limit on the break frequency, we find the spectral index of the optically thick component to be $\gtrsim 0.18$. The optically thin component has a negative spectral index $\sim -0.7$. This suggests that the jet spectrum possibly curves to become flat somewhere in the IR, implying the break to be close to the IR/optical bands \citep[e.g.,][]{gandhi11}. 

\subsubsection{Optical polarimetry}

There are several potential sources of optical polarization in XRBs. Firstly, the disk emission may be polarized up to $12\%$ depending on the inclination, and the polarization angle can vary with the opacity of the atmosphere, switching by $\sim 90^\circ$ between the optically thin and thick cases \citep{chandrasekhar60, sunyaev85}. Secondly, synchrotron emission from the jet can be polarized up to $70-75~\%$ depending on the slope of the electron distribution and the level of ordering of the magnetic field lines in the jet \citep{rybicki79}.  The resulting linear polarization in the total emission of the XRB will, however, strongly depend on the fractional contribution of the jet to the optical flux, which varies depending on the source and on the time of the observation. In addition, synchrotron emission from a hot flow can also give a small polarization (up to a few per cent) if the magnetic field in the hot flow has an ordered component \citep{veledina13}. Although the expected polarization from various mechanisms is well understood theoretically, it is often difficult to disentangle the individual components. 

Our polarization measurements of \source{} with FORS2/VLT suggest an interesting scenario. Of the three epochs of observations (Figure \ref{fig:polevo}), the first two were acquired during the intermediate state, while the last one was obtained during the hard state. From the previous analyses (Sections \ref{sec:sedcmddis} and \ref{sec:corrdis}), we found that the jet spectrum possibly extends to the optical wavebands, and could therefore be responsible for the observed linear polarization. In accordance with this scenario, the polarization spectrum is flat, imitating the flat OIR flux spectrum of the jet. On the other hand, if the polarization was caused by electron scatting in the disc, it should have increased toward bluer frequencies. Such low levels of OIR polarization from the jet are a strong indication of tangled magnetic fields at the base of the jet, as is typically observed in most XRBs \citep[see e.g.,][]{russelld18a,baglio20}. 

The third epoch, during the excursion to the hard state, saw a dramatic change in polarization. While $P$ could not be constrained in the $B$ and \uvv{} bands, it increased to $1.03\pm0.06~\%$ and $0.63\pm0.06~\%$ in the $R$ and $I$ bands, respectively. Moreover, $\theta$ rotated by $\sim100^\circ$ compared to the previous epochs. It is interesting to note that this third observation was made just before ($<1$ day) the short X-ray flare on MJD 60150, after which the source transitioned to the soft state. In BHXBs, transitions from the hard intermediate state to the soft intermediate state are often accompanied by X-ray brightening and/or discrete ejection of ballistic jets. In this circumstance, compression of the magnetic field lines in shocks within the jet might occur, generating partially ordered transverse magnetic fields \citep[as observed in the case of the BHXB V404 Cyg;][]{shahbaz16}. This can explain the roughly orthogonal rotation of $\theta$ in the third epoch as compared to the first two epochs. Moreover, a discrete ejection would also imply an optically thin synchrotron emission, i.e., a red spectrum, with stronger polarization at longer wavelengths, as observed (Figure \ref{fig:polevo}). Incidentally, the rotation of $\theta$ by $90^\circ$ during the transition has only been observed at radio wavelengths \citep[e.g.,][]{curran14,curran15}. Therefore, \source{} could be the first XRB to display a similar swing in $\theta$ in the OIR wavebands associated with the state change (and in this case, a flare), which could potentially be due to the quenching of the compact jet and the ejection of the ballistic jet during the hard-to-soft transition. 

\subsection{Reflares}

Outbursts of XRBs, as well as those of accreting white dwarfs in dwarf novae, often exhibit rebrightening episodes during and/or after the end of the main outburst \citep[e.g.,][]{chen93, chen97}. Depending on a rebrightening's amplitude, shape, duration, and position in the lightcurve it can be classified as a glitch, a reflare, or a mini-outburst \citep[see][for a detailed classification scheme]{zhanggb19}. The rebrightenings in \source{} appear to start before the main outburst returns to quiescence, and their peak brightness remains less than $70\%$ of the peak brightness of the main outburst, so we classify them as reflares (Figure \ref{fig:lc}). 

The high amplitude and short duration of the reflares are one of the most unusual aspects of this source, seldom seen in other NSXBs. The peak brightness of these reflares in both the X-ray and optical/UV bands rises by $\sim 2$ orders of magnitude above the quiescent levels (Figure \ref{fig:lc}). Not only is the amplitude high, the source also reaches the peak flux levels rapidly in $\sim 2-4$ days, matching the rapid rise of the outburst onset. Some sources like Swift J1858.6$-0814$ show numerous short-timescale reflares, but they are of low amplitude \citep[$\lesssim 1$ mag][]{rhodes24}. Then there are other sources like IGR J$00291+5934$, which show large-amplitude reflares, but they are of long duration and few \citep{lewis10}. The high amplitudes and short duty cycles of the reflares in \source{} make it difficult to explain theoretically \citep[see][for reflares in SAX J1808.4--3658 which are similar to \source{}]{patruno16}.  

Reflares that occur near in time, or just after, the main outburst pose a challenge for the DIM as at this stage the disk is almost depleted of matter. The propeller effect has been invoked in this regard that can halt the disk from emptying itself near the end of an outburst, thereby maintaining a matter reservoir for further accretion events \citep{hartman11, patruno16}. It has also been suggested that a hot non-standard, or trapped, disk can exist towards the end of the outburst that accretes matter at low levels \citep{patruno16,zhanggb19, cuneo20}. However, a strong propeller regime is disfavored by \citet{rout25}, as strong magnetic fields would result in very little accretion onto the NS surface. Since the radius of the inner disk often reaches the last stable orbit, a weak propeller or a trapped disk scenario is considered more likely \citep{rout25}. In this configuration, the inner parts of the disk deviate from the standard thin-disk solution and low levels of episodic accretion persist on the NS surface \citep{patruno16}. 

An alternative explanation from the DIM is that reflares could be caused by multiple reflections of the heating and cooling fronts whenever they encounter a density gradient \citep{lasota01}. This process was invoked to explain the reflares in the BHXB Swift J1910.2--0546 \citep{saikia23}. However, the CMDs for \source{} (Figure \ref{fig:cmd}) imply that the disk temperature always remained in the hot branch ($> 7000$ K), even at \gp{} $\sim 22$ mag, which is inconsistent with the idea of traversing heating/cooling fronts. Irradiation of the outer accretion disk could keep the disk in the hot branch and prevent quiescence from setting in \citep{king98}. The high values of the OIR spectral index ($0.75 \lesssim \alpha \lesssim 1.4$) during the reflare peaks suggest sustained irradiation of the outer disk as a viable mechanism. The repeating nature of the reflares can then arise due to heating of the companion star that results in an enhanced mass transfer in a quasi-periodic manner, or similar ``echoes" from an earlier enhanced accretion event \citep{augusteijn93}.

Outflows have also been proposed as an explanation for reflares in XRBs. The OIR reflares in BHXBs are sometimes caused by jet emission after the transition to the hard state \citep[e.g.,][]{buxton04, kalemci13}. However, this is unlikely to be the case for \source{} as the emission at the reflaring stage is dominated by the irradiated disk. A jet, if present, would be fainter than the disk to be directly detected. 

\section{Summary} \label{sec:summary}

The NS atoll \source{} went into an outburst in July 2023. We observed the source in radio, IR, optical, UV, and X-ray wavebands. The main findings of our study are as follows:
\begin{itemize}
    \item The optical rise was detected about 4-12 days before the X-ray rise. Such a delay is consistent with a truncated disk scenario where the heating front first reaches the disk edge, which then moves towards the compact object on viscous time scales to produce X-rays.
    
    \item After reaching the peak of the outburst, in the intermediate and hard state, the presence of jet is supported by radio and IR detection. The jet synchrotron emission probably also dominates the optical wavebands as evidenced from the evolution of optical color and SEDs.  
 
    \item The powerlaw indices of the correlations between the X-ray and UVOIR fluxes are high ($\beta \sim0.6 - 0.9$). Steep correlations are detected at longer wavelengths (i.e., \rp{}, \ip{}, and \zs{} bands), suggesting a possible contribution of the jet.
    
    \item Optical polarization measured with the VLT showed low levels of polarization ($\lesssim 1\%$) in the first two epochs. The third epoch, just before an X-ray flare, saw a rotation of the polarization angle by $\sim 100^\circ$ in the $R$ and $I$ bands. This could be related to an ejection of material associated with the flare.
    
    \item The main outburst is followed by a series of high-amplitude rapid reflares in the optical, UV, and X-ray wavebands. The reflare SEDs suggest a reprocessed disk emission in the optical/UV wavebands, but their color temperatures are inconsistent with traveling heating/cooling fronts.  
\end{itemize}

\begin{acknowledgments}
The authors acknowledge constructive feedback from the referee. This research is based on work supported by Tamkeen under the NYU Abu Dhabi Research Institute grant CASS. This research has made use of data and/or software provided by the High Energy Astrophysics Science Archive Research Center (HEASARC), which is a service of the Astrophysics Science Division at NASA/GSFC. This work made use of data supplied by the UK Swift Science Data Centre at the University of Leicester. This research is based on observations collected at the European Southern Observatory under ESO programme 111.24K2. MCB acknowledges support from the INAF-Astrofit fellowship. TMD acknowledges support by the Spanish Ministry of Science via the Plan de Generacion de conocimiento PID2021-124879NB-I00. JH acknowledges support through NASA grant 80NSSC23K1659. MAP acknowledges support through the Ramón y Cajal grant RYC2022-035388-I, funded by MCIU/AEI/10.13039/501100011033 and FSE$+$. NM acknowledges financial support through ASI-INAF agreement 2017-14-H.0 (PI: T. Belloni). JvdE acknowledges a Warwick Astrophysics prize post-doctoral fellowship made possible thanks to a generous philanthropic donation, and was supported by funding from the European Union's Horizon Europe research and innovation programme under the Marie Sk\l{}odowska-Curie grant agreement No 101148693 (MeerSHOCKS) for part of this work. Mangum, and Mark Durre.
\end{acknowledgments}

\facilities{NICER, Swift(XRT and UVOT), LCO, VLT, REM, VLA, MeerKAT}
\software{Astropy \citep{astropy}, Scipy \citep{scipy}, Numpy \citep{numpy}, Matplotlib \citep{matplotlib}}

\newpage

\appendix

\section{Optical Polarimetry with VLT} \label{sec:optpol}

To determine the linear polarization of \source, we applied the algorithm described by \cite{baglio20} and references therein. This method begins by calculating the parameter $S(\Phi)$ for each HWP angle:

\begin{equation}
S(\Phi)=\left( \frac{f^{o}(\Phi)/f^e(\Phi)}{f^o_u(\Phi)/f^e_u(\Phi)}-1\right)/\left( \frac{f^{o}(\Phi)/f^e(\Phi)}{f^o_u(\Phi)/f^e_u(\Phi)}+1\right),
\end{equation}

where $f^o(\Phi)$ and $f^e(\Phi)$ are the ordinary and extraordinary fluxes of \source, respectively, and $f^o_u(\Phi)$ and $f^e_u(\Phi)$ are the corresponding values for an unpolarized standard star in the field. This parameter is related to the target's polarization degree ($P$) and polarization angle ($\theta$) by the equation:

\begin{equation}\label{eq_cos}
S(\Phi) = P\, \cos 2(\theta - \Phi).
\end{equation}

Therefore, a fit of the $S$ parameter with Equation~\ref{eq_cos} will give an estimate of the linear $P$ and $\theta$ for the target. To increase the significance of the fit, we considered 6 reference field stars in each epoch. Under the simple hypothesis that the field stars are intrinsically unpolarized, this method gives as a result a linear polarization for the target that is already corrected for the low instrumental effects. Moreover, if the field stars are polarized due to interstellar dust, this method should, in principle, automatically correct for interstellar polarization along the line of sight.  

Following \citet{baglio20}, to evaluate $P$ and $\theta$ we maximized the Gaussian likelihood function using an optimization algorithm \citep[e.g., the Nelder–Mead algorithm;][]{gao12} and integrated the posterior probability density of our model parameters using a Markov Chain Monte Carlo (MCMC) algorithm \citep{Hogg&Foreman2018} based on the ``affine-invariant Hamiltonian'' algorithm \citep{Foreman-Mackeyetal2013}. The chains were initiated from small Gaussian distributions centered on the best-fit values. We discarded the first third of each chain as the ``burn-in phase'' and ensured that a stationary distribution was reached \citep{Sharma2017}. The quality of the fit was assessed as described in \cite{Lucy2016}. The values for $P$ and $\theta$, along with their $1\sigma$ uncertainties, correspond to the 0.16, 0.50, and 0.84 quantiles of the posterior distribution of the parameters. In the case of non-detections, the 99.97\% percentile of the posterior distribution of the parameter $P$ was used to estimate an upper limit. The value of $\theta$ derived using this method was further adjusted based on observations of the polarized standard star BD-12 5133 (observed during the night of 2023 July 22 with the same setup as that of our observations), with known and documented polarization angles in all FORS2 bands \citep{Fossati2007}. The average correction applied was negligible, remaining under $2^{\circ}$ across all bands and epochs.

\section{VLA Campaign} \label{sec:vla}

The apparent temporal variability observed with MeerKAT was initially believed to be due to the discrete jet ejections common to both BH \citep[e.g.,][]{han92_AKH,hjellming95,atetarenko17_AKH,bright20_AKH} and (Z-source) NS LMXBs \citep[e.g.,][although they are significantly less than BHs]{fomalont01_AKH,spencer13_AKH}. To confirm the existence of the jets, we were approved for director's discretionary time observations (Project Code: VLA/23B-302) with the Very Large Array (VLA). At the time, the VLA was in its most extended A-configuration, resulting in a factor of $\sim$\,5 improvement in angular resolution compared to MeerKAT (at L-band). We acquired nine VLA observations: eight 8-bit L-band observations ($\sim$\,1.5 GHz, $\sim$\,1 GHz bandwidth) between 2023 September 15 (MJD 60202) and 2023 October 9 (MJD 60226) and a single 3-bit C-band observation ($\sim$\,6 GHz, $\sim$\,4 GHz bandwidth) at the same epoch as with the first L-band observation. During the last two (L-band) observations, the VLA moved into a hybrid configuration (A $\rightarrow$ D), significantly increasing the background flux density and, thus, noise levels; the increase in background was due to D-configuration's sensitivity to large-scale extended emission. As a result, we do not use the last two VLA observations. 

We processed the VLA data with the \textsc{casa} pipeline \citep[v6.4;][]{casa22}. For each observation, we passed the data through twice; after the first round of calibration, we inspected the visibilities, manually flagging any residual corrupted data and re-running the pipeline. Following calibration, we adopted the \textsc{oxkat} imaging workflow, producing one continuum image per VLA epoch.

The high-resolution VLA observations revealed our initial interpretations were incorrect; i.e., the radio morphology did not result from jet ejecta. Instead, we discovered three moderately variable ($\sim$\,100 $\mu$Jy with $\sim$\,30\% excess variances) but unassociated point sources within $\sim$\,10$^{\prime\prime}$ of MAXI J1807$+$132  (see Figure \ref{fig:radio}). The apparent motion was due to MeerKAT's inadequate angular resolution combined with the variability of the three field sources, leading to confusion about the morphological evolution of MAXI J1807$+132$. We used the \textsc{casa} task \textsc{imfit} to measure the positions of each source in each VLA observation, enforcing that each source adopts the shape of the synthesized beam. Below, we present the inverse-variance weighted average positions and errors using ${>}\,4\sigma$ detections (with the VLA):

\begin{enumerate}[label=Source \arabic*:, leftmargin=*]
    \item 18h08m08.528s $+13^\circ15^\prime07.77^{\prime\prime}\,(\pm0.08^{\prime\prime})$
    \item 18h08m08.040s $+13^\circ15^\prime05.87^{\prime\prime}\,(\pm0.05^{\prime\prime})$
    \item 18h08m07.373s $+13^\circ15^\prime09.13^{\prime\prime}\,(\pm0.17^{\prime\prime})$
\end{enumerate}

In addition to the astrometric errors from \textsc{imfit} \citep[which follow,][]{Condon97_AKH,Condon98_AKH}, we added, in quadrature, a systematic astrometric error equal to 10\% of the (major-axis) FWHM of the synthesized beam\footnote{This systematic error follows the VLA recommendation: \ \url{https://science.nrao.edu/facilities/vla/docs/manuals/oss/performance/positional-accuracy}}. We investigated the accuracy of our errors utilizing a nearby (18h08m22.455s $+13^\circ17^\prime26.129^{\prime\prime}$) $\sim$\,3\,mJy field source, comparing our measured position to the positions fit using the VLA Sky Survey (VLASS) quick-look images\footnote{The quick-look image server can be found here, \  \url{http://cutouts.cirada.ca/}}. The offsets of the check source in our images (compared to VLASS) are ${<}\,0.05^{\prime\prime}$, and have a uniform distribution of offset position angles (i.e., the direction of the offset, measured East of North). Given that our adopted astrometric systematic error is ${>}\,0.1^{\prime\prime}$ for all epochs, our estimated astrometric errors are reasonable, if not overly conservative. Future radio monitoring campaigns of MAXI J1807$+$132 need to compensate for the associated field sources; the MeerKAT S-band receivers or the VLA in A-configuration have sufficient angular resolution to mitigate source confusion.

\section{Supplementary Figure: Fits to long-term lightcurve }

\begin{figure}[h]
    \centering
    \includegraphics[scale=0.7]{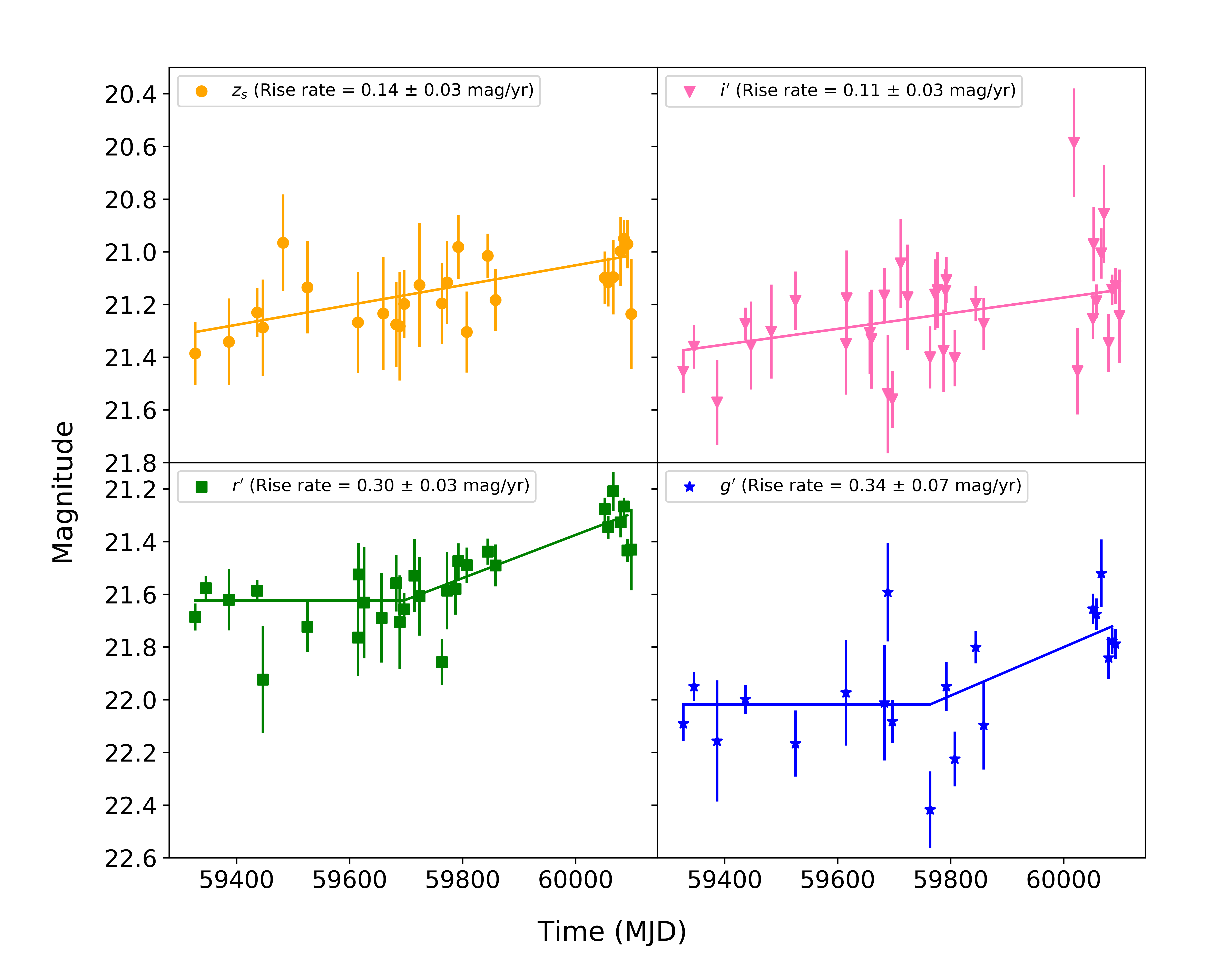}
    \caption{Fits to the long-term quiescent lightcurves of \source{} with the slow rise model. The color and symbols for different panels are the same as in Figure \ref{fig:lcquiesc}. The legends also mention the rise rates for the different filters.}
    \label{fig:slowrise}
\end{figure}

\newpage

\bibliography{references}{}

\begin{thebibliography}{}
\expandafter\ifx\csname natexlab\endcsname\relax\def\natexlab#1{#1}\fi
\providecommand{\url}[1]{\href{#1}{#1}}
\providecommand{\dodoi}[1]{doi:~\href{http://doi.org/#1}{\nolinkurl{#1}}}
\providecommand{\doeprint}[1]{\href{http://ascl.net/#1}{\nolinkurl{http://ascl.net/#1}}}
\providecommand{\doarXiv}[1]{\href{https://arxiv.org/abs/#1}{\nolinkurl{https://arxiv.org/abs/#1}}}

\bibitem[{A.~C. {Albayati} {et~al.}(2021){Albayati}, {Altamirano}, {Jaisawal}, {Bult}, {Rapisarda}, {Mancuso}, {G{\"u}ver}, {Arzoumanian}, {Chakrabarty}, {Chenevez}, {Court}, {Gendreau}, {Guillot}, {Keek}, {Malacaria}, \& {Strohmayer}}]{albayati21}
{Albayati}, A.~C., {Altamirano}, D., {Jaisawal}, G.~K., {et~al.} 2021, \bibinfo{title}{{Discovery of thermonuclear Type-I X-ray bursts from the X-ray binary MAXI J1807+132},} \mnras, 501, 261, \dodoi{10.1093/mnras/staa3657}

\bibitem[{I. {Appenzeller} {et~al.}(1998){Appenzeller}, {Fricke}, {F{\"u}rtig}, {G{\"a}ssler}, {H{\"a}fner}, {Harke}, {Hess}, {Hummel}, {J{\"u}rgens}, {Kudritzki}, {Mantel}, {Meisl}, {Muschielok}, {Nicklas}, {Rupprecht}, {Seifert}, {Stahl}, {Szeifert}, \& {Tarantik}}]{appenzeller98}
{Appenzeller}, I., {Fricke}, K., {F{\"u}rtig}, W., {et~al.} 1998, \bibinfo{title}{{Successful commissioning of FORS1 - the first optical instrument on the VLT.},} The Messenger, 94, 1

\bibitem[{M. {Armas Padilla} {et~al.}(2013){Armas Padilla}, {Degenaar}, {Russell}, \& {Wijnands}}]{armaspadilla13}
{Armas Padilla}, M., {Degenaar}, N., {Russell}, D.~M., \& {Wijnands}, R. 2013, \bibinfo{title}{{Multiwavelength spectral evolution during the 2011 outburst of the very faint X-ray transient Swift J1357.2-0933},} \mnras, 428, 3083, \dodoi{10.1093/mnras/sts255}

\bibitem[{ {Astropy Collaboration} {et~al.}(2022){Astropy Collaboration}, {Price-Whelan}, {Lim}, {Earl}, {Starkman}, {Bradley}, {Shupe}, {Patil}, {Corrales}, {Brasseur}, {N{\"o}the}, {Donath}, {Tollerud}, {Morris}, {Ginsburg}, {Vaher}, {Weaver}, {Tocknell}, {Jamieson}, {van Kerkwijk}, {Robitaille}, {Merry}, {Bachetti}, {G{\"u}nther}, {Aldcroft}, {Alvarado-Montes}, {Archibald}, {B{\'o}di}, {Bapat}, {Barentsen}, {Baz{\'a}n}, {Biswas}, {Boquien}, {Burke}, {Cara}, {Cara}, {Conroy}, {Conseil}, {Craig}, {Cross}, {Cruz}, {D'Eugenio}, {Dencheva}, {Devillepoix}, {Dietrich}, {Eigenbrot}, {Erben}, {Ferreira}, {Foreman-Mackey}, {Fox}, {Freij}, {Garg}, {Geda}, {Glattly}, {Gondhalekar}, {Gordon}, {Grant}, {Greenfield}, {Groener}, {Guest}, {Gurovich}, {Handberg}, {Hart}, {Hatfield-Dodds}, {Homeier}, {Hosseinzadeh}, {Jenness}, {Jones}, {Joseph}, {Kalmbach}, {Karamehmetoglu}, {Ka{\l}uszy{\'n}ski}, {Kelley}, {Kern}, {Kerzendorf}, {Koch}, {Kulumani}, {Lee}, {Ly}, {Ma}, {MacBride}, {Maljaars}, {Muna}, {Murphy}, {Norman},
  {O'Steen}, {Oman}, {Pacifici}, {Pascual}, {Pascual-Granado}, {Patil}, {Perren}, {Pickering}, {Rastogi}, {Roulston}, {Ryan}, {Rykoff}, {Sabater}, {Sakurikar}, {Salgado}, {Sanghi}, {Saunders}, {Savchenko}, {Schwardt}, {Seifert-Eckert}, {Shih}, {Jain}, {Shukla}, {Sick}, {Simpson}, {Singanamalla}, {Singer}, {Singhal}, {Sinha}, {Sip{\H{o}}cz}, {Spitler}, {Stansby}, {Streicher}, {{\v{S}}umak}, {Swinbank}, {Taranu}, {Tewary}, {Tremblay}, {de Val-Borro}, {Van Kooten}, {Vasovi{\'c}}, {Verma}, {de Miranda Cardoso}, {Williams}, {Wilson}, {Winkel}, {Wood-Vasey}, {Xue}, {Yoachim}, {Zhang}, {Zonca}, \& {Astropy Project Contributors}}]{astropy}
{Astropy Collaboration}, {Price-Whelan}, A.~M., {Lim}, P.~L., {et~al.} 2022, \bibinfo{title}{{The Astropy Project: Sustaining and Growing a Community-oriented Open-source Project and the Latest Major Release (v5.0) of the Core Package},} \apj, 935, 167, \dodoi{10.3847/1538-4357/ac7c74}

\bibitem[{T. {Augusteijn} {et~al.}(1993){Augusteijn}, {Kuulkers}, \& {Shaham}}]{augusteijn93}
{Augusteijn}, T., {Kuulkers}, E., \& {Shaham}, J. 1993, \bibinfo{title}{{``Glitches'' in soft X-ray transients : echoes of the main burst ?},} \aap, 279, L13

\bibitem[{M.~C. {Baglio} {et~al.}(2020){Baglio}, {Russell}, {Crespi}, {Covino}, {Johar}, {Homan}, {Bramich}, {Saikia}, {Campana}, {D'Avanzo}, {Fender}, {Goldoni}, {Goodwin}, {Lewis}, {Masetti}, {Miraval Zanon}, {Motta}, {Mu{\~n}oz-Darias}, \& {Shahbaz}}]{baglio20}
{Baglio}, M.~C., {Russell}, D.~M., {Crespi}, S., {et~al.} 2020, \bibinfo{title}{{Probing Jet Launching in Neutron Star X-Ray Binaries: The Variable and Polarized Jet of SAX J1808.4-3658},} \apj, 905, 87, \dodoi{10.3847/1538-4357/abc685}

\bibitem[{M.~C. {Baglio} {et~al.}(2022){Baglio}, {Saikia}, {Russell}, {Homan}, {Waterval}, {Bramich}, {Campana}, {Lewis}, {Eijnden}, {Alabarta}, {Covino}, {D'Avanzo}, {Goldoni}, {Masetti}, \& {Mu{\~n}oz-Darias}}]{baglio22}
{Baglio}, M.~C., {Saikia}, P., {Russell}, D.~M., {et~al.} 2022, \bibinfo{title}{{A Misfired Outburst in the Neutron Star X-Ray Binary Centaurus X-4},} \apj, 930, 20, \dodoi{10.3847/1538-4357/ac63ad}

\bibitem[{F. {Bernardini} {et~al.}(2016){Bernardini}, {Russell}, {Shaw}, {Lewis}, {Charles}, {Koljonen}, {Lasota}, \& {Casares}}]{bernardini16a}
{Bernardini}, F., {Russell}, D.~M., {Shaw}, A.~W., {et~al.} 2016, \bibinfo{title}{{Events leading up to the 2015 June Outburst of V404 Cyg},} \apjl, 818, L5, \dodoi{10.3847/2041-8205/818/1/L5}

\bibitem[{R.~D. {Blandford} \& A. {K{\"o}nigl}(1979){Blandford} \& {K{\"o}nigl}}]{blandford79}
{Blandford}, R.~D., \& {K{\"o}nigl}, A. 1979, \bibinfo{title}{{Relativistic jets as compact radio sources.},} \apj, 232, 34, \dodoi{10.1086/157262}

\bibitem[{D.~M. {Bramich} \& W. {Freudling}(2012){Bramich} \& {Freudling}}]{bramich12}
{Bramich}, D.~M., \& {Freudling}, W. 2012, \bibinfo{title}{{Systematic trends in Sloan Digital Sky Survey photometric data},} \mnras, 424, 1584, \dodoi{10.1111/j.1365-2966.2012.21385.x}

\bibitem[{J.~S. Bright {et~al.}(2020)Bright {et~al.}}]{bright20_AKH}
Bright, J.~S., {et~al.} 2020, \bibinfo{title}{{An extremely powerful long-lived superluminal ejection from the black hole MAXI J1820+070},} Nature Astron., 4, 697, \dodoi{10.1038/s41550-020-1023-5}

\bibitem[{C. {Brocksopp} {et~al.}(2001){Brocksopp}, {Jonker}, {Fender}, {Groot}, {van der Klis}, \& {Tingay}}]{brocksopp01}
{Brocksopp}, C., {Jonker}, P.~G., {Fender}, R.~P., {et~al.} 2001, \bibinfo{title}{{The 1997 hard-state outburst of the X-ray transient GS 1354-64/BW Cir},} \mnras, 323, 517, \dodoi{10.1046/j.1365-8711.2001.04193.x}

\bibitem[{D.~N. {Burrows} {et~al.}(2005){Burrows}, {Hill}, {Nousek}, {Kennea}, {Wells}, {Osborne}, {Abbey}, {Beardmore}, {Mukerjee}, {Short}, {Chincarini}, {Campana}, {Citterio}, {Moretti}, {Pagani}, {Tagliaferri}, {Giommi}, {Capalbi}, {Tamburelli}, {Angelini}, {Cusumano}, {Br{\"a}uninger}, {Burkert}, \& {Hartner}}]{burrows05}
{Burrows}, D.~N., {Hill}, J.~E., {Nousek}, J.~A., {et~al.} 2005, \bibinfo{title}{{The Swift X-Ray Telescope},} \ssr, 120, 165, \dodoi{10.1007/s11214-005-5097-2}

\bibitem[{M.~M. {Buxton} \& C.~D. {Bailyn}(2004){Buxton} \& {Bailyn}}]{buxton04}
{Buxton}, M.~M., \& {Bailyn}, C.~D. 2004, \bibinfo{title}{{The 2002 Outburst of the Black Hole X-Ray Binary 4U 1543-47: Optical and Infrared Light Curves},} \apj, 615, 880, \dodoi{10.1086/424503}

\bibitem[{J.~A. {Cardelli} {et~al.}(1989){Cardelli}, {Clayton}, \& {Mathis}}]{cardelli89}
{Cardelli}, J.~A., {Clayton}, G.~C., \& {Mathis}, J.~S. 1989, \bibinfo{title}{{The Relationship between Infrared, Optical, and Ultraviolet Extinction},} \apj, 345, 245, \dodoi{10.1086/167900}

\bibitem[{F. {Carotenuto} {et~al.}(2021){Carotenuto}, {Corbel}, {Tremou}, {Russell}, {Tzioumis}, {Fender}, {Woudt}, {Motta}, {Miller-Jones}, {Chauhan}, {Tetarenko}, {Sivakoff}, {Heywood}, {Horesh}, {van der Horst}, {Koerding}, \& {Mooley}}]{carotenuto21}
{Carotenuto}, F., {Corbel}, S., {Tremou}, E., {et~al.} 2021, \bibinfo{title}{{The black hole transient MAXI J1348-630: evolution of the compact and transient jets during its 2019/2020 outburst},} \mnras, 504, 444, \dodoi{10.1093/mnras/stab864}

\bibitem[{ {CASA Team} {et~al.}(2022){CASA Team}, {Bean}, {Bhatnagar}, {Castro}, {Donovan Meyer}, {Emonts}, {Garcia}, {Garwood}, {Golap}, {Gonzalez Villalba}, {Harris}, {Hayashi}, {Hoskins}, {Hsieh}, {Jagannathan}, {Kawasaki}, {Keimpema}, {Kettenis}, {Lopez}, {Marvil}, {Masters}, {McNichols}, {Mehringer}, {Miel}, {Moellenbrock}, {Montesino}, {Nakazato}, {Ott}, {Petry}, {Pokorny}, {Raba}, {Rau}, {Schiebel}, {Schweighart}, {Sekhar}, {Shimada}, {Small}, {Steeb}, {Sugimoto}, {Suoranta}, {Tsutsumi}, {van Bemmel}, {Verkouter}, {Wells}, {Xiong}, {Szomoru}, {Griffith}, {Glendenning}, \& {Kern}}]{casa22}
{CASA Team}, {Bean}, B., {Bhatnagar}, S., {et~al.} 2022, \bibinfo{title}{{CASA, the Common Astronomy Software Applications for Radio Astronomy},} \pasp, 134, 114501, \dodoi{10.1088/1538-3873/ac9642}

\bibitem[{S. {Chandrasekhar}(1960){Chandrasekhar}}]{chandrasekhar60}
{Chandrasekhar}, S. 1960, {Radiative transfer}

\bibitem[{W. {Chen} {et~al.}(1993){Chen}, {Livio}, \& {Gehrels}}]{chen93}
{Chen}, W., {Livio}, M., \& {Gehrels}, N. 1993, \bibinfo{title}{{The Secondary Maxima in Black Hole X-Ray Nova Light Curves: Clues toward a Complete Picture},} \apjl, 408, L5, \dodoi{10.1086/186817}

\bibitem[{W. {Chen} {et~al.}(1997){Chen}, {Shrader}, \& {Livio}}]{chen97}
{Chen}, W., {Shrader}, C.~R., \& {Livio}, M. 1997, \bibinfo{title}{{The Properties of X-Ray and Optical Light Curves of X-Ray Novae},} \apj, 491, 312, \dodoi{10.1086/304921}

\bibitem[{J.~J. {Condon}(1997){Condon}}]{Condon97_AKH}
{Condon}, J.~J. 1997, \bibinfo{title}{{Errors in Elliptical Gaussian Fits},} \pasp, 109, 166, \dodoi{10.1086/133871}

\bibitem[{J.~J. {Condon} {et~al.}(1998){Condon}, {Cotton}, {Greisen}, {Yin}, {Perley}, {Taylor}, \& {Broderick}}]{Condon98_AKH}
{Condon}, J.~J., {Cotton}, W.~D., {Greisen}, E.~W., {et~al.} 1998, \bibinfo{title}{{The NRAO VLA Sky Survey},} \aj, 115, 1693, \dodoi{10.1086/300337}

\bibitem[{S. {Corbel} \& R.~P. {Fender}(2002){Corbel} \& {Fender}}]{corbel02}
{Corbel}, S., \& {Fender}, R.~P. 2002, \bibinfo{title}{{Near-Infrared Synchrotron Emission from the Compact Jet of GX 339-4},} \apjl, 573, L35, \dodoi{10.1086/341870}

\bibitem[{V.~A. {C{\'u}neo} {et~al.}(2020){C{\'u}neo}, {Alabarta}, {Zhang}, {Altamirano}, {M{\'e}ndez}, {Armas Padilla}, {Remillard}, {Homan}, {Steiner}, {Combi}, {Mu{\~n}oz-Darias}, {Gendreau}, {Arzoumanian}, {Stevens}, {Loewenstein}, {Tombesi}, {Bult}, {Fabian}, {Buisson}, {Neilsen}, \& {Basak}}]{cuneo20}
{C{\'u}neo}, V.~A., {Alabarta}, K., {Zhang}, L., {et~al.} 2020, \bibinfo{title}{{A NICER look at the state transitions of the black hole candidate MAXI J1535-571 during its reflares},} \mnras, 496, 1001, \dodoi{10.1093/mnras/staa1606}

\bibitem[{C. {Cunningham}(1976){Cunningham}}]{cunningham76}
{Cunningham}, C. 1976, \bibinfo{title}{{Returning radiation in accretion disks around black holes.},} \apj, 208, 534, \dodoi{10.1086/154636}

\bibitem[{P.~A. {Curran} {et~al.}(2014){Curran}, {Coriat}, {Miller-Jones}, {Armstrong}, {Edwards}, {Sivakoff}, {Woudt}, {Altamirano}, {Belloni}, {Corbel}, {Fender}, {K{\"o}rding}, {Krimm}, {Markoff}, {Migliari}, {Russell}, {Stevens}, \& {Tzioumis}}]{curran14}
{Curran}, P.~A., {Coriat}, M., {Miller-Jones}, J.~C.~A., {et~al.} 2014, \bibinfo{title}{{The evolving polarized jet of black hole candidate Swift J1745-26},} \mnras, 437, 3265, \dodoi{10.1093/mnras/stt2125}

\bibitem[{P.~A. {Curran} {et~al.}(2015){Curran}, {Miller-Jones}, {Rushton}, {Pawar}, {Anderson}, {Altamirano}, {Krimm}, {Broderick}, {Belloni}, {Fender}, {K{\"o}rding}, {Maitra}, {Markoff}, {Migliari}, {Rumsey}, {Rupen}, {Russell}, {Russell}, {Sarazin}, {Sivakoff}, {Soria}, {Tetarenko}, {Titterington}, \& {Tudose}}]{curran15}
{Curran}, P.~A., {Miller-Jones}, J.~C.~A., {Rushton}, A.~P., {et~al.} 2015, \bibinfo{title}{{Radio polarimetry as a probe of unresolved jets: the 2013 outburst of XTE J1908+094},} \mnras, 451, 3975, \dodoi{10.1093/mnras/stv1252}

\bibitem[{N. {Degenaar} {et~al.}(2014){Degenaar}, {Maitra}, {Cackett}, {Reynolds}, {Miller}, {Reis}, {King}, {G{\"u}ltekin}, {Bailyn}, {Buxton}, {MacDonald}, {Fabian}, {Fox}, \& {Rykoff}}]{degenaar14}
{Degenaar}, N., {Maitra}, D., {Cackett}, E.~M., {et~al.} 2014, \bibinfo{title}{{Multi-wavelength Coverage of State Transitions in the New Black Hole X-Ray Binary Swift J1910.2-0546},} \apj, 784, 122, \dodoi{10.1088/0004-637X/784/2/122}

\bibitem[{M. {D{\'\i}az Trigo} {et~al.}(2018){D{\'\i}az Trigo}, {Altamirano}, {Din{\c{c}}er}, {Miller-Jones}, {Russell}, {Sanna}, {Bailyn}, {Lewis}, {Migliari}, \& {Rahoui}}]{diaztrigo18}
{D{\'\i}az Trigo}, M., {Altamirano}, D., {Din{\c{c}}er}, T., {et~al.} 2018, \bibinfo{title}{{The evolving jet spectrum of the neutron star X-ray binary Aql X-1 in transitional states during its 2016 outburst},} \aap, 616, A23, \dodoi{10.1051/0004-6361/201832693}

\bibitem[{G. {Dubus} {et~al.}(2001){Dubus}, {Hameury}, \& {Lasota}}]{dubus01}
{Dubus}, G., {Hameury}, J.~M., \& {Lasota}, J.~P. 2001, \bibinfo{title}{{The disc instability model for X-ray transients: Evidence for truncation and irradiation},} \aap, 373, 251, \dodoi{10.1051/0004-6361:20010632}

\bibitem[{P.~A. {Evans} {et~al.}(2007){Evans}, {Beardmore}, {Page}, {Tyler}, {Osborne}, {Goad}, {O'Brien}, {Vetere}, {Racusin}, {Morris}, {Burrows}, {Capalbi}, {Perri}, {Gehrels}, \& {Romano}}]{evans07}
{Evans}, P.~A., {Beardmore}, A.~P., {Page}, K.~L., {et~al.} 2007, \bibinfo{title}{{An online repository of Swift/XRT light curves of {\ensuremath{\gamma}}-ray bursts},} \aap, 469, 379, \dodoi{10.1051/0004-6361:20077530}

\bibitem[{P.~A. {Evans} {et~al.}(2009){Evans}, {Beardmore}, {Page}, {Osborne}, {O'Brien}, {Willingale}, {Starling}, {Burrows}, {Godet}, {Vetere}, {Racusin}, {Goad}, {Wiersema}, {Angelini}, {Capalbi}, {Chincarini}, {Gehrels}, {Kennea}, {Margutti}, {Morris}, {Mountford}, {Pagani}, {Perri}, {Romano}, \& {Tanvir}}]{evans09}
{Evans}, P.~A., {Beardmore}, A.~P., {Page}, K.~L., {et~al.} 2009, \bibinfo{title}{{Methods and results of an automatic analysis of a complete sample of Swift-XRT observations of GRBs},} \mnras, 397, 1177, \dodoi{10.1111/j.1365-2966.2009.14913.x}

\bibitem[{H. {Falcke} {et~al.}(2004){Falcke}, {K{\"o}rding}, \& {Markoff}}]{falcke04}
{Falcke}, H., {K{\"o}rding}, E., \& {Markoff}, S. 2004, \bibinfo{title}{{A scheme to unify low-power accreting black holes. Jet-dominated accretion flows and the radio/X-ray correlation},} \aap, 414, 895, \dodoi{10.1051/0004-6361:20031683}

\bibitem[{R. {Fender} \& E. {Gallo}(2014){Fender} \& {Gallo}}]{fender14}
{Fender}, R., \& {Gallo}, E. 2014, \bibinfo{title}{{An Overview of Jets and Outflows in Stellar Mass Black Holes},} \ssr, 183, 323, \dodoi{10.1007/s11214-014-0069-z}

\bibitem[{R. {Fender} \& T. {Mu{\~n}oz-Darias}(2016){Fender} \& {Mu{\~n}oz-Darias}}]{fender16}
{Fender}, R., \& {Mu{\~n}oz-Darias}, T. 2016, in Lecture Notes in Physics, Berlin Springer Verlag, ed. F.~{Haardt}, V.~{Gorini}, U.~{Moschella}, A.~{Treves}, \& M.~{Colpi}, Vol. 905, 65, \dodoi{10.1007/978-3-319-19416-5_3}

\bibitem[{R. {Fender} {et~al.}(2016){Fender}, {Woudt}, {Corbel}, {Coriat}, {Daigne}, {Falcke}, {Girard}, {Heywood}, {Horesh}, {Horrell}, {Jonker}, {Joseph}, {Kamble}, {Knigge}, {K{\"o}rding}, {Kotze}, {Kouveliotou}, {Lynch}, {Maccarone}, {Meintjes}, {Migliari}, {Murphy}, {Nagayama}, {Nelemans}, {Nicholson}, {O'Brien}, {Oodendaal}, {Oozeer}, {Osborne}, {P{\'e}rez-Torres}, {Ratcliffe}, {Ribeiro}, {Rol}, {Rushton}, {Scaife}, {Schurch}, {Sivakoff}, {Staley}, {Steeghs}, {Stewart}, {Swinbank}, {Vergani}, {Warner}, {Wiersema}, {Armstrong}, {Groot}, {McBride}, {Miller-Jones}, {Mooley}, {Stappers}, {Wijers}, {Bietenholz}, {Blyth}, {B{\"o}ttcher}, {Buckley}, {Charles}, {Chomiuk}, {Coppejans}, {de Blok}, {van der Heyden}, {van der Horst}, \& {van Soelen}}]{tkat17_AKH}
{Fender}, R., {Woudt}, P.~A., {Corbel}, S., {et~al.} 2016, in MeerKAT Science: On the Pathway to the SKA, 13, \dodoi{10.22323/1.277.0013}

\bibitem[{R.~P. {Fender} {et~al.}(2004){Fender}, {Belloni}, \& {Gallo}}]{fender04b}
{Fender}, R.~P., {Belloni}, T.~M., \& {Gallo}, E. 2004, \bibinfo{title}{{Towards a unified model for black hole X-ray binary jets},} \mnras, 355, 1105, \dodoi{10.1111/j.1365-2966.2004.08384.x}

\bibitem[{R.~P. {Fender} {et~al.}(2001){Fender}, {Hjellming}, {Tilanus}, {Pooley}, {Deane}, {Ogley}, \& {Spencer}}]{fender01a}
{Fender}, R.~P., {Hjellming}, R.~M., {Tilanus}, R.~P.~J., {et~al.} 2001, \bibinfo{title}{{Spectral evidence for a powerful compact jet from XTE J1118+480},} \mnras, 322, L23, \dodoi{10.1046/j.1365-8711.2001.04362.x}

\bibitem[{R.~P. {Fender} {et~al.}(2009){Fender}, {Homan}, \& {Belloni}}]{fender09}
{Fender}, R.~P., {Homan}, J., \& {Belloni}, T.~M. 2009, \bibinfo{title}{{Jets from black hole X-ray binaries: testing, refining and extending empirical models for the coupling to X-rays},} \mnras, 396, 1370, \dodoi{10.1111/j.1365-2966.2009.14841.x}

\bibitem[{R.~P. {Fender} \& E. {Kuulkers}(2001){Fender} \& {Kuulkers}}]{fender01b}
{Fender}, R.~P., \& {Kuulkers}, E. 2001, \bibinfo{title}{{On the peak radio and X-ray emission from neutron star and black hole candidate X-ray transients},} \mnras, 324, 923, \dodoi{10.1046/j.1365-8711.2001.04345.x}

\bibitem[{S. {Fijma} {et~al.}(2023){Fijma}, {van den Eijnden}, {Degenaar}, {Russell}, \& {Miller-Jones}}]{fijma23}
{Fijma}, S., {van den Eijnden}, J., {Degenaar}, N., {Russell}, T.~D., \& {Miller-Jones}, J.~C.~A. 2023, \bibinfo{title}{{Evaluating the jet/accretion coupling of Aql X-1: probing the contribution of accretion flow spectral components},} \mnras, 521, 4490, \dodoi{10.1093/mnras/stad548}

\bibitem[{D.~R. {Foight} {et~al.}(2016){Foight}, {G{\"u}ver}, {{\"O}zel}, \& {Slane}}]{foight16}
{Foight}, D.~R., {G{\"u}ver}, T., {{\"O}zel}, F., \& {Slane}, P.~O. 2016, \bibinfo{title}{{Probing X-Ray Absorption and Optical Extinction in the Interstellar Medium Using Chandra Observations of Supernova Remnants},} \apj, 826, 66, \dodoi{10.3847/0004-637X/826/1/66}

\bibitem[{E.~B. {Fomalont} {et~al.}(2001){Fomalont}, {Geldzahler}, \& {Bradshaw}}]{fomalont01_AKH}
{Fomalont}, E.~B., {Geldzahler}, B.~J., \& {Bradshaw}, C.~F. 2001, \bibinfo{title}{{Scorpius X-1: The Evolution and Nature of the Twin Compact Radio Lobes},} \apj, 558, 283, \dodoi{10.1086/322479}

\bibitem[{D. {Foreman-Mackey} {et~al.}(2013){Foreman-Mackey}, {Hogg}, {Lang}, \& {Goodman}}]{Foreman-Mackeyetal2013}
{Foreman-Mackey}, D., {Hogg}, D.~W., {Lang}, D., \& {Goodman}, J. 2013, \bibinfo{title}{{emcee: The MCMC Hammer},} \pasp, 125, 306, \dodoi{10.1086/670067}

\bibitem[{L. {Fossati} {et~al.}(2007){Fossati}, {Bagnulo}, {Mason}, \& {Landi Degl'Innocenti}}]{Fossati2007}
{Fossati}, L., {Bagnulo}, S., {Mason}, E., \& {Landi Degl'Innocenti}, E. 2007, in Astronomical Society of the Pacific Conference Series, Vol. 364, The Future of Photometric, Spectrophotometric and Polarimetric Standardization, ed. C.~{Sterken}, 503

\bibitem[{J. {Frank} {et~al.}(2002){Frank}, {King}, \& {Raine}}]{frank02}
{Frank}, J., {King}, A., \& {Raine}, D.~J. 2002, {Accretion Power in Astrophysics: Third Edition} ({Cambridge University Press, Cambridge, UK})

\bibitem[{E. {Gallo} {et~al.}(2018){Gallo}, {Degenaar}, \& {van den Eijnden}}]{gallo18}
{Gallo}, E., {Degenaar}, N., \& {van den Eijnden}, J. 2018, \bibinfo{title}{{Hard state neutron star and black hole X-ray binaries in the radio:X-ray luminosity plane},} \mnras, 478, L132, \dodoi{10.1093/mnrasl/sly083}

\bibitem[{E. {Gallo} {et~al.}(2012){Gallo}, {Miller}, \& {Fender}}]{gallo12}
{Gallo}, E., {Miller}, B.~P., \& {Fender}, R. 2012, \bibinfo{title}{{Assessing luminosity correlations via cluster analysis: evidence for dual tracks in the radio/X-ray domain of black hole X-ray binaries},} \mnras, 423, 590, \dodoi{10.1111/j.1365-2966.2012.20899.x}

\bibitem[{P. {Gandhi} {et~al.}(2011){Gandhi}, {Blain}, {Russell}, {Casella}, {Malzac}, {Corbel}, {D'Avanzo}, {Lewis}, {Markoff}, {Cadolle Bel}, {Goldoni}, {Wachter}, {Khangulyan}, \& {Mainzer}}]{gandhi11}
{Gandhi}, P., {Blain}, A.~W., {Russell}, D.~M., {et~al.} 2011, \bibinfo{title}{{A Variable Mid-infrared Synchrotron Break Associated with the Compact Jet in GX 339-4},} \apjl, 740, L13, \dodoi{10.1088/2041-8205/740/1/L13}

\bibitem[{F. {Gao} \& L. {Han}(2012){Gao} \& {Han}}]{gao12}
{Gao}, F., \& {Han}, L. 2012, \bibinfo{title}{{Implementing the Nelder-Mead simplex algorithm with adaptive parameters},} Computational Optimization and Applications, \dodoi{10.1007/s10589-010-9329-3}

\bibitem[{K.~V.~S. {Gasealahwe} {et~al.}(2024){Gasealahwe}, {Monageng}, {Fender}, {Woudt}, {Hughes}, {Motta}, {van den Eijnden}, {Saikia}, \& {Tremou}}]{gasealahwe24}
{Gasealahwe}, K.~V.~S., {Monageng}, I.~M., {Fender}, R.~P., {et~al.} 2024, \bibinfo{title}{{Radio observations of the 2022 outburst of the transitional Z-Atoll source XTE J1701-462},} \mnras, 533, 1800, \dodoi{10.1093/mnras/stae1875}

\bibitem[{K.~C. {Gendreau} {et~al.}(2016){Gendreau}, {Arzoumanian}, {Adkins}, {Albert}, {Anders}, {Aylward}, {Baker}, {Balsamo}, {Bamford}, {Benegalrao}, {Berry}, {Bhalwani}, {Black}, {Blaurock}, {Bronke}, {Brown}, {Budinoff}, {Cantwell}, {Cazeau}, {Chen}, {Clement}, {Colangelo}, {Coleman}, {Coopersmith}, {Dehaven}, {Doty}, {Egan}, {Enoto}, {Fan}, {Ferro}, {Foster}, {Galassi}, {Gallo}, {Green}, {Grosh}, {Ha}, {Hasouneh}, {Heefner}, {Hestnes}, {Hoge}, {Jacobs}, {J{\o}rgensen}, {Kaiser}, {Kellogg}, {Kenyon}, {Koenecke}, {Kozon}, {LaMarr}, {Lambertson}, {Larson}, {Lentine}, {Lewis}, {Lilly}, {Liu}, {Malonis}, {Manthripragada}, {Markwardt}, {Matonak}, {Mcginnis}, {Miller}, {Mitchell}, {Mitchell}, {Mohammed}, {Monroe}, {Montt de Garcia}, {Mul{\'e}}, {Nagao}, {Ngo}, {Norris}, {Norwood}, {Novotka}, {Okajima}, {Olsen}, {Onyeachu}, {Orosco}, {Peterson}, {Pevear}, {Pham}, {Pollard}, {Pope}, {Powers}, {Powers}, {Price}, {Prigozhin}, {Ramirez}, {Reid}, {Remillard}, {Rogstad}, {Rosecrans}, {Rowe}, {Sager}, {Sanders},
  {Savadkin}, {Saylor}, {Schaeffer}, {Schweiss}, {Semper}, {Serlemitsos}, {Shackelford}, {Soong}, {Struebel}, {Vezie}, {Villasenor}, {Winternitz}, {Wofford}, {Wright}, {Yang}, \& {Yu}}]{gendreau16}
{Gendreau}, K.~C., {Arzoumanian}, Z., {Adkins}, P.~W., {et~al.} 2016, in Society of Photo-Optical Instrumentation Engineers (SPIE) Conference Series, Vol. 9905, Space Telescopes and Instrumentation 2016: Ultraviolet to Gamma Ray, ed. J.-W.~A. {den Herder}, T.~{Takahashi}, \& M.~{Bautz}, 99051H, \dodoi{10.1117/12.2231304}

\bibitem[{A.~J. {Goodwin} {et~al.}(2020){Goodwin}, {Russell}, {Galloway}, {Baglio}, {Parikh}, {Buckley}, {Homan}, {Bramich}, {in't Zand}, {Heinke}, {Kotze}, {de Martino}, {Papitto}, {Lewis}, \& {Wijnands}}]{goodwin20}
{Goodwin}, A.~J., {Russell}, D.~M., {Galloway}, D.~K., {et~al.} 2020, \bibinfo{title}{{Enhanced optical activity 12 d before X-ray activity, and a 4 d X-ray delay during outburst rise, in a low-mass X-ray binary},} \mnras, 498, 3429, \dodoi{10.1093/mnras/staa2588}

\bibitem[{N.~V. {Gusinskaia} {et~al.}(2020){Gusinskaia}, {Hessels}, {Degenaar}, {Deller}, {Miller-Jones}, {Archibald}, {Heinke}, {Mold{\'o}n}, {Patruno}, {Tomsick}, \& {Wijnands}}]{gusinskaia20}
{Gusinskaia}, N.~V., {Hessels}, J.~W.~T., {Degenaar}, N., {et~al.} 2020, \bibinfo{title}{{Quasi-simultaneous radio and X-ray observations of Aql X-1 : probing low luminosities},} \mnras, 492, 2858, \dodoi{10.1093/mnras/stz3420}

\bibitem[{J.~M. {Hameury}(2020){Hameury}}]{hameury20}
{Hameury}, J.~M. 2020, \bibinfo{title}{{A review of the disc instability model for dwarf novae, soft X-ray transients and related objects},} Advances in Space Research, 66, 1004, \dodoi{10.1016/j.asr.2019.10.022}

\bibitem[{J.~M. {Hameury} {et~al.}(1997){Hameury}, {Lasota}, {McClintock}, \& {Narayan}}]{hameury97}
{Hameury}, J.~M., {Lasota}, J.~P., {McClintock}, J.~E., \& {Narayan}, R. 1997, \bibinfo{title}{{Advection-dominated Flows around Black Holes and the X-Ray Delay in the Outburst of GRO J1655-40},} \apj, 489, 234, \dodoi{10.1086/304780}

\bibitem[{X. {Han} \& R.~M. {Hjellming}(1992){Han} \& {Hjellming}}]{han92_AKH}
{Han}, X., \& {Hjellming}, R.~M. 1992, \bibinfo{title}{{Radio Observations of the 1989 Transient Event in V404 Cygni (= GS 2023+338)},} \apj, 400, 304, \dodoi{10.1086/171996}

\bibitem[{C.~R. Harris {et~al.}(2020)Harris, Millman, van~der Walt, Gommers, Virtanen, Cournapeau, Wieser, Taylor, Berg, Smith, Kern, Picus, Hoyer, van Kerkwijk, Brett, Haldane, del R{\'{i}}o, Wiebe, Peterson, G{\'{e}}rard-Marchant, Sheppard, Reddy, Weckesser, Abbasi, Gohlke, \& Oliphant}]{numpy}
Harris, C.~R., Millman, K.~J., van~der Walt, S.~J., {et~al.} 2020, \bibinfo{title}{Array programming with {NumPy},} Nature, 585, 357, \dodoi{10.1038/s41586-020-2649-2}

\bibitem[{J.~M. {Hartman} {et~al.}(2011){Hartman}, {Galloway}, \& {Chakrabarty}}]{hartman11}
{Hartman}, J.~M., {Galloway}, D.~K., \& {Chakrabarty}, D. 2011, \bibinfo{title}{{A Double Outburst from IGR J00291+5934: Implications for Accretion Disk Instability Theory},} \apj, 726, 26, \dodoi{10.1088/0004-637X/726/1/26}

\bibitem[{G. {Hasinger} \& M. {van der Klis}(1989){Hasinger} \& {van der Klis}}]{hasinger89}
{Hasinger}, G., \& {van der Klis}, M. 1989, \bibinfo{title}{{Two patterns of correlated X-ray timing and spectral behaviour in low-mass X-ray binaries.},} \aap, 225, 79

\bibitem[{I. {Heywood}(2020){Heywood}}]{oxkat20_AKH}
{Heywood}, I. 2020, \bibinfo{title}{{oxkat: Semi-automated imaging of MeerKAT observations},}, Astrophysics Source Code Library, record ascl:2009.003 \doeprint{2009.003}

\bibitem[{I. {Heywood} {et~al.}(2022){Heywood}, {Jarvis}, {Hale}, {Whittam}, {Bester}, {Hugo}, {Kenyon}, {Prescott}, {Smirnov}, {Tasse}, {Afonso}, {Best}, {Collier}, {Deane}, {Frank}, {Hardcastle}, {Knowles}, {Maddox}, {Murphy}, {Prandoni}, {Randriamampandry}, {Santos}, {Sekhar}, {Tabatabaei}, {Taylor}, \& {Thorat}}]{mightee20_AKH}
{Heywood}, I., {Jarvis}, M.~J., {Hale}, C.~L., {et~al.} 2022, \bibinfo{title}{{MIGHTEE: total intensity radio continuum imaging and the COSMOS/XMM-LSS Early Science fields},} \mnras, 509, 2150, \dodoi{10.1093/mnras/stab3021}

\bibitem[{R.~M. {Hjellming} \& M.~P. {Rupen}(1995){Hjellming} \& {Rupen}}]{hjellming95}
{Hjellming}, R.~M., \& {Rupen}, M.~P. 1995, \bibinfo{title}{{Episodic ejection of relativistic jets by the X-ray transient GRO J1655 - 40},} \nat, 375, 464, \dodoi{10.1038/375464a0}

\bibitem[{R.~M. {Hjellming} {et~al.}(1990){Hjellming}, {Stewart}, {White}, {Strom}, {Lewin}, {Hertz}, {Wood}, {Norris}, {Mitsuda}, {Penninx}, \& {van Paradijs}}]{hjellming90}
{Hjellming}, R.~M., {Stewart}, R.~T., {White}, G.~L., {et~al.} 1990, \bibinfo{title}{{Radio and X-Ray States in the X-Ray Binary Scorpius X-1},} \apj, 365, 681, \dodoi{10.1086/169522}

\bibitem[{D.~W. {Hogg} \& D. {Foreman-Mackey}(2018){Hogg} \& {Foreman-Mackey}}]{Hogg&Foreman2018}
{Hogg}, D.~W., \& {Foreman-Mackey}, D. 2018, \bibinfo{title}{{Data Analysis Recipes: Using Markov Chain Monte Carlo},} \apjs, 236, 11, \dodoi{10.3847/1538-4365/aab76e}

\bibitem[{J. {Homan} {et~al.}(2005){Homan}, {Buxton}, {Markoff}, {Bailyn}, {Nespoli}, \& {Belloni}}]{homan05a}
{Homan}, J., {Buxton}, M., {Markoff}, S., {et~al.} 2005, \bibinfo{title}{{Multiwavelength Observations of the 2002 Outburst of GX 339-4: Two Patterns of X-Ray-Optical/Near-Infrared Behavior},} \apj, 624, 295, \dodoi{10.1086/428722}

\bibitem[{J. {Homan} {et~al.}(2016){Homan}, {Neilsen}, {Allen}, {Chakrabarty}, {Fender}, {Fridriksson}, {Remillard}, \& {Schulz}}]{homan16}
{Homan}, J., {Neilsen}, J., {Allen}, J.~L., {et~al.} 2016, \bibinfo{title}{{Evidence for Simultaneous Jets and Disk Winds in Luminous Low-mass X-Ray Binaries},} \apjl, 830, L5, \dodoi{10.3847/2041-8205/830/1/L5}

\bibitem[{J.~D. Hunter(2007)Hunter}]{matplotlib}
Hunter, J.~D. 2007, \bibinfo{title}{Matplotlib: A 2D graphics environment,} Computing in Science \& Engineering, 9, 90, \dodoi{10.1109/MCSE.2007.55}

\bibitem[{R.~I. {Hynes}(2005){Hynes}}]{hynes05}
{Hynes}, R.~I. 2005, \bibinfo{title}{{The Optical and Ultraviolet Spectral Energy Distributions of Short-Period Black Hole X-Ray Transients in Outburst},} \apj, 623, 1026, \dodoi{10.1086/428445}

\bibitem[{G. {Illiano} {et~al.}(2023){Illiano}, {Papitto}, {Ambrosino}, {Zanon}, \& {Sanna}}]{atelilliano23}
{Illiano}, G., {Papitto}, A., {Ambrosino}, F., {Zanon}, A.~M., \& {Sanna}, A. 2023, \bibinfo{title}{{NICER confirms the onset of a new outburst from MAXI J1807+132},} The Astronomer's Telegram, 16125, 1

\bibitem[{F. {Jim{\'e}nez-Ibarra} {et~al.}(2019){Jim{\'e}nez-Ibarra}, {Mu{\~n}oz-Darias}, {Armas Padilla}, {Russell}, {Casares}, {Torres}, {Mata S{\'a}nchez}, {Jonker}, \& {Lewis}}]{jiminezibarra19}
{Jim{\'e}nez-Ibarra}, F., {Mu{\~n}oz-Darias}, T., {Armas Padilla}, M., {et~al.} 2019, \bibinfo{title}{{The complex evolution of the X-ray binary transient MAXI J1807+132 along the decay of its discovery outburst},} \mnras, 484, 2078, \dodoi{10.1093/mnras/sty3457}

\bibitem[{C. {John} {et~al.}(2024){John}, {De}, {Lucchini}, {Behar}, {Kara}, {MacLeod}, {Panagiotou}, \& {Wang}}]{john24}
{John}, C., {De}, K., {Lucchini}, M., {et~al.} 2024, \bibinfo{title}{{Correlated mid-infrared and X-ray outbursts in black hole X-ray binaries: A new route to discovery in infrared surveys},} arXiv e-prints, arXiv:2406.17866, \dodoi{10.48550/arXiv.2406.17866}

\bibitem[{J. {Jonas} \&  {MeerKAT Team}(2016){Jonas} \& {MeerKAT Team}}]{Jonas16_AKH}
{Jonas}, J., \& {MeerKAT Team}. 2016, in MeerKAT Science: On the Pathway to the SKA, 1, \dodoi{10.22323/1.277.0001}

\bibitem[{E. {Kalemci} {et~al.}(2013){Kalemci}, {Din{\c{c}}er}, {Tomsick}, {Buxton}, {Bailyn}, \& {Chun}}]{kalemci13}
{Kalemci}, E., {Din{\c{c}}er}, T., {Tomsick}, J.~A., {et~al.} 2013, \bibinfo{title}{{Complete Multiwavelength Evolution of Galactic Black Hole Transients during Outburst Decay. I. Conditions for ``Compact'' Jet Formation},} \apj, 779, 95, \dodoi{10.1088/0004-637X/779/2/95}

\bibitem[{A.~R. {King} \& H. {Ritter}(1998){King} \& {Ritter}}]{king98}
{King}, A.~R., \& {Ritter}, H. 1998, \bibinfo{title}{{The light curves of soft X-ray transients},} \mnras, 293, L42, \dodoi{10.1046/j.1365-8711.1998.01295.x}

\bibitem[{K.~I.~I. {Koljonen} {et~al.}(2016){Koljonen}, {Russell}, {Corral-Santana}, {Armas Padilla}, {Mu{\~n}oz-Darias}, {Lewis}, {Coriat}, \& {Bauer}}]{koljonen16}
{Koljonen}, K.~I.~I., {Russell}, D.~M., {Corral-Santana}, J.~M., {et~al.} 2016, \bibinfo{title}{{A `high-hard' outburst of the black hole X-ray binary GS 1354-64},} \mnras, 460, 942, \dodoi{10.1093/mnras/stw1007}

\bibitem[{J.-P. {Lasota}(2001){Lasota}}]{lasota01}
{Lasota}, J.-P. 2001, \bibinfo{title}{{The disc instability model of dwarf novae and low-mass X-ray binary transients},} \nar, 45, 449, \dodoi{10.1016/S1387-6473(01)00112-9}

\bibitem[{F. {Lewis} {et~al.}(2008){Lewis}, {Russell}, {Fender}, {Roche}, \& {Clark}}]{lewis08a}
{Lewis}, F., {Russell}, D.~M., {Fender}, R.~P., {Roche}, P., \& {Clark}, J.~S. 2008, \bibinfo{title}{{Continued Monitoring of LMXBs with the Faulkes Telescopes},} arXiv e-prints, arXiv:0811.2336, \dodoi{10.48550/arXiv.0811.2336}

\bibitem[{F. {Lewis} {et~al.}(2010){Lewis}, {Russell}, {Jonker}, {Linares}, {Tudose}, {Roche}, {Clark}, {Torres}, {Maitra}, {Bassa}, {Steeghs}, {Patruno}, {Migliari}, {Wijnands}, {Nelemans}, {Kewley}, {Stroud}, {Modjaz}, {Bloom}, {Blake}, \& {Starr}}]{lewis10}
{Lewis}, F., {Russell}, D.~M., {Jonker}, P.~G., {et~al.} 2010, \bibinfo{title}{{The double-peaked 2008 outburst of the accreting milli-second X-ray pulsar, IGR J00291+5934},} \aap, 517, A72, \dodoi{10.1051/0004-6361/201014382}

\bibitem[{M. {Livio} \& J.~E. {Pringle}(1992){Livio} \& {Pringle}}]{livio92}
{Livio}, M., \& {Pringle}, J.~E. 1992, \bibinfo{title}{{Dwarf nova outbursts - the ultraviolet delay and the effect of a weakly magnetized white dwarf.},} \mnras, 259, 23P, \dodoi{10.1093/mnras/259.1.23P}

\bibitem[{L.~B. {Lucy}(2016){Lucy}}]{Lucy2016}
{Lucy}, L.~B. 2016, \bibinfo{title}{{Frequentist tests for Bayesian models},} \aap, 588, A19, \dodoi{10.1051/0004-6361/201527709}

\bibitem[{D. {Maitra} \& C.~D. {Bailyn}(2008){Maitra} \& {Bailyn}}]{maitra08}
{Maitra}, D., \& {Bailyn}, C.~D. 2008, \bibinfo{title}{{Outburst Morphology in the Soft X-Ray Transient Aquila X-1},} \apj, 688, 537, \dodoi{10.1086/592029}

\bibitem[{K. {Menou} {et~al.}(1999){Menou}, {Hameury}, \& {Stehle}}]{menou99}
{Menou}, K., {Hameury}, J.-M., \& {Stehle}, R. 1999, \bibinfo{title}{{Structure and properties of transition fronts in accretion discs},} \mnras, 305, 79, \dodoi{10.1046/j.1365-8711.1999.02396.x}

\bibitem[{F. {Meyer} \& E. {Meyer-Hofmeister}(1994){Meyer} \& {Meyer-Hofmeister}}]{meyer94}
{Meyer}, F., \& {Meyer-Hofmeister}, E. 1994, \bibinfo{title}{{Accretion disk evaporation by a coronal siphon flow.},} \aap, 288, 175

\bibitem[{S. {Migliari} \& R.~P. {Fender}(2006){Migliari} \& {Fender}}]{migliari06a}
{Migliari}, S., \& {Fender}, R.~P. 2006, \bibinfo{title}{{Jets in neutron star X-ray binaries: a comparison with black holes},} \mnras, 366, 79, \dodoi{10.1111/j.1365-2966.2005.09777.x}

\bibitem[{S. {Migliari} {et~al.}(2004){Migliari}, {Fender}, {Rupen}, {Wachter}, {Jonker}, {Homan}, \& {van der Klis}}]{migliari04}
{Migliari}, S., {Fender}, R.~P., {Rupen}, M., {et~al.} 2004, \bibinfo{title}{{Radio detections of the neutron star X-ray binaries 4U 1820 - 30 and Ser X-1 in soft X-ray states},} \mnras, 351, 186, \dodoi{10.1111/j.1365-2966.2004.07768.x}

\bibitem[{A. {Patruno} {et~al.}(2016){Patruno}, {Maitra}, {Curran}, {D'Angelo}, {Fridriksson}, {Russell}, {Middleton}, \& {Wijnands}}]{patruno16}
{Patruno}, A., {Maitra}, D., {Curran}, P.~A., {et~al.} 2016, \bibinfo{title}{{The Reflares and Outburst Evolution in the Accreting Millisecond Pulsar SAX J1808.4-3658: A Disk Truncated Near Co-Rotation?},} \apj, 817, 100, \dodoi{10.3847/0004-637X/817/2/100}

\bibitem[{W. {Penninx} {et~al.}(1988){Penninx}, {Lewin}, {Zijlstra}, {Mitsuda}, \& {van Paradijs}}]{penninx88}
{Penninx}, W., {Lewin}, W. H.~G., {Zijlstra}, A.~A., {Mitsuda}, K., \& {van Paradijs}, J. 1988, \bibinfo{title}{{A connection between the X-ray spectral branches and the radio brightness in GX17 + 2},} \nat, 336, 146, \dodoi{10.1038/336146a0}

\bibitem[{L. {Rhodes} {et~al.}(2024){Rhodes}, {Russell}, {Saikia}, {Alabarta}, {van den Eijnden}, {Knight}, {Baglio}, \& {Lewis}}]{rhodes24}
{Rhodes}, L., {Russell}, D.~M., {Saikia}, P., {et~al.} 2024, \bibinfo{title}{{Long term optical variations in Swift J1858.6-0814: evidence for ablation and comparisons to radio properties},} \mnras, \dodoi{10.1093/mnras/stae2755}

\bibitem[{P.~W.~A. {Roming} {et~al.}(2005){Roming}, {Kennedy}, {Mason}, {Nousek}, {Ahr}, {Bingham}, {Broos}, {Carter}, {Hancock}, {Huckle}, {Hunsberger}, {Kawakami}, {Killough}, {Koch}, {McLelland}, {Smith}, {Smith}, {Soto}, {Boyd}, {Breeveld}, {Holland}, {Ivanushkina}, {Pryzby}, {Still}, \& {Stock}}]{roming05}
{Roming}, P. W.~A., {Kennedy}, T.~E., {Mason}, K.~O., {et~al.} 2005, \bibinfo{title}{{The Swift Ultra-Violet/Optical Telescope},} \ssr, 120, 95, \dodoi{10.1007/s11214-005-5095-4}

\bibitem[{S.~K. {Rout} {et~al.}(2025){Rout}, {Mu{\~n}oz-Darias}, {Homan}, {Armas Padilla}, {Russell}, {Alabarta}, \& {Saikia}}]{rout25}
{Rout}, S.~K., {Mu{\~n}oz-Darias}, T., {Homan}, J., {et~al.} 2025, \bibinfo{title}{{Evolution of the Accretion Disk and Corona during the Outburst of the Neutron Star Transient MAXI J1807+132},} \apj, 978, 12, \dodoi{10.3847/1538-4357/ad919f}

\bibitem[{S.~K. {Rout} {et~al.}(2021){Rout}, {Vadawale}, {Aarthy}, {Ganesh}, {Joshi}, {Roy}, {Misra}, \& {Yadav}}]{rout21b}
{Rout}, S.~K., {Vadawale}, S.~V., {Aarthy}, E., {et~al.} 2021, \bibinfo{title}{{Multi-wavelength view of the galactic black-hole binary GRS 1716-249},} Journal of Astrophysics and Astronomy, 42, 39, \dodoi{10.1007/s12036-021-09696-5}

\bibitem[{D.~M. {Russell}(2018){Russell}}]{russelld18a}
{Russell}, D.~M. 2018, \bibinfo{title}{{Optical/Infrared Polarised Emission in X-ray Binaries},} Galaxies, 6, 3, \dodoi{10.3390/galaxies6010003}

\bibitem[{D.~M. {Russell} {et~al.}(2006){Russell}, {Fender}, {Hynes}, {Brocksopp}, {Homan}, {Jonker}, \& {Buxton}}]{russelld06}
{Russell}, D.~M., {Fender}, R.~P., {Hynes}, R.~I., {et~al.} 2006, \bibinfo{title}{{Global optical/infrared-X-ray correlations in X-ray binaries: quantifying disc and jet contributions},} \mnras, 371, 1334, \dodoi{10.1111/j.1365-2966.2006.10756.x}

\bibitem[{D.~M. {Russell} {et~al.}(2007){Russell}, {Fender}, \& {Jonker}}]{russelld07}
{Russell}, D.~M., {Fender}, R.~P., \& {Jonker}, P.~G. 2007, \bibinfo{title}{{Evidence for a jet contribution to the optical/infrared light of neutron star X-ray binaries},} \mnras, 379, 1108, \dodoi{10.1111/j.1365-2966.2007.12008.x}

\bibitem[{D.~M. {Russell} {et~al.}(2010){Russell}, {Lewis}, {Roche}, {Clark}, {Breedt}, \& {Fender}}]{russelld10a}
{Russell}, D.~M., {Lewis}, F., {Roche}, P., {et~al.} 2010, \bibinfo{title}{{A long-term optical-X-ray correlation in 4U 1957+11},} \mnras, 402, 2671, \dodoi{10.1111/j.1365-2966.2009.16098.x}

\bibitem[{D.~M. {Russell} {et~al.}(2011){Russell}, {Maitra}, {Dunn}, \& {Fender}}]{russelld11}
{Russell}, D.~M., {Maitra}, D., {Dunn}, R.~J.~H., \& {Fender}, R.~P. 2011, \bibinfo{title}{{A tool to separate optical/infrared disc and jet emission in X-ray transient outbursts: the colour-magnitude diagrams of XTE J1550-564},} \mnras, 416, 2311, \dodoi{10.1111/j.1365-2966.2011.19204.x}

\bibitem[{D.~M. {Russell} {et~al.}(2018){Russell}, {Qasim}, {Bernardini}, {Plotkin}, {Lewis}, {Koljonen}, \& {Yang}}]{russelld18b}
{Russell}, D.~M., {Qasim}, A.~A., {Bernardini}, F., {et~al.} 2018, \bibinfo{title}{{Optical Precursors to Black Hole X-Ray Binary Outbursts: An Evolving Synchrotron Jet Spectrum in Swift J1357.2-0933},} \apj, 852, 90, \dodoi{10.3847/1538-4357/aa9d8c}

\bibitem[{D.~M. {Russell} {et~al.}(2013{\natexlab{a}}){Russell}, {Markoff}, {Casella}, {Cantrell}, {Chatterjee}, {Fender}, {Gallo}, {Gandhi}, {Homan}, {Maitra}, {Miller-Jones}, {O'Brien}, \& {Shahbaz}}]{russelld13a}
{Russell}, D.~M., {Markoff}, S., {Casella}, P., {et~al.} 2013{\natexlab{a}}, \bibinfo{title}{{Jet spectral breaks in black hole X-ray binaries},} \mnras, 429, 815, \dodoi{10.1093/mnras/sts377}

\bibitem[{D.~M. {Russell} {et~al.}(2013{\natexlab{b}}){Russell}, {Russell}, {Miller-Jones}, {O'Brien}, {Soria}, {Sivakoff}, {Slaven-Blair}, {Lewis}, {Markoff}, {Homan}, {Altamirano}, {Curran}, {Rupen}, {Belloni}, {Cadolle Bel}, {Casella}, {Corbel}, {Dhawan}, {Fender}, {Gallo}, {Gandhi}, {Heinz}, {K{\"o}rding}, {Krimm}, {Maitra}, {Migliari}, {Remillard}, {Sarazin}, {Shahbaz}, \& {Tudose}}]{russelld13b}
{Russell}, D.~M., {Russell}, T.~D., {Miller-Jones}, J.~C.~A., {et~al.} 2013{\natexlab{b}}, \bibinfo{title}{{An Evolving Compact Jet in the Black Hole X-Ray Binary MAXI J1836-194},} \apjl, 768, L35, \dodoi{10.1088/2041-8205/768/2/L35}

\bibitem[{D.~M. {Russell} {et~al.}(2019){Russell}, {Bramich}, {Lewis}, {AlMannaei}, {Al Qaissieh}, {Al Qasim}, {Al Yazeedi}, {Baglio}, {Bernardini}, {Elgalad}, {Gabuya}, {Lasota}, {Palado}, {Roche}, {Shivkumar}, {Udrescu}, \& {Zhang}}]{russelld19}
{Russell}, D.~M., {Bramich}, D.~M., {Lewis}, F., {et~al.} 2019, \bibinfo{title}{{Optical precursors to X‑ray binary outbursts},} Astronomische Nachrichten, 340, 278, \dodoi{10.1002/asna.201913610}

\bibitem[{T.~D. {Russell} {et~al.}(2014){Russell}, {Soria}, {Miller-Jones}, {Curran}, {Markoff}, {Russell}, \& {Sivakoff}}]{russellt14}
{Russell}, T.~D., {Soria}, R., {Miller-Jones}, J.~C.~A., {et~al.} 2014, \bibinfo{title}{{The accretion-ejection coupling in the black hole candidate X-ray binary MAXI J1836-194},} \mnras, 439, 1390, \dodoi{10.1093/mnras/stt2498}

\bibitem[{T.~D. {Russell} {et~al.}(2021){Russell}, {Degenaar}, {van den Eijnden}, {Del Santo}, {Segreto}, {Altamirano}, {Beri}, {D{\'\i}az Trigo}, \& {Miller-Jones}}]{russellt21}
{Russell}, T.~D., {Degenaar}, N., {van den Eijnden}, J., {et~al.} 2021, \bibinfo{title}{{The evolving radio jet from the neutron star X-ray binary 4U 1820-30},} \mnras, 508, L6, \dodoi{10.1093/mnrasl/slab087}

\bibitem[{G.~B. {Rybicki} \& A.~P. {Lightman}(1979){Rybicki} \& {Lightman}}]{rybicki79}
{Rybicki}, G.~B., \& {Lightman}, A.~P. 1979, {Radiative processes in astrophysics}

\bibitem[{P. {Saikia} {et~al.}(2023{\natexlab{a}}){Saikia}, {Russell}, {Alabarta}, {Baglio}, {Bramich}, {Homan}, {Russell}, \& {Lewis}}]{atelsaikia23a}
{Saikia}, P., {Russell}, D.~M., {Alabarta}, K., {et~al.} 2023{\natexlab{a}}, \bibinfo{title}{{XB-NEWS detects a new outburst from the X-ray transient MAXI J1807+132},} The Astronomer's Telegram, 16119, 1

\bibitem[{P. {Saikia} {et~al.}(2023{\natexlab{b}}){Saikia}, {Russell}, {Pirbhoy}, {Baglio}, {Bramich}, {Alabarta}, {Lewis}, \& {Charles}}]{saikia23}
{Saikia}, P., {Russell}, D.~M., {Pirbhoy}, S.~F., {et~al.} 2023{\natexlab{b}}, \bibinfo{title}{{Seven Reflares, a Mini Outburst, and an Outburst: High-amplitude Optical Variations in the Black Hole X-Ray Binary Swift J1910.2-0546},} \apj, 949, 104, \dodoi{10.3847/1538-4357/acc8cc}

\bibitem[{P. {Saikia} {et~al.}(2023{\natexlab{c}}){Saikia}, {Homan}, {Alabarta}, {Russell}, {Baglio}, {Bramich}, {Rout}, {Russell}, \& {Lewis}}]{atelsaikia23b}
{Saikia}, P., {Homan}, J., {Alabarta}, K., {et~al.} 2023{\natexlab{c}}, \bibinfo{title}{{An optical/X-ray re-flare from the X-ray transient MAXI J1807+132},} The Astronomer's Telegram, 16185, 1

\bibitem[{T. {Shahbaz} {et~al.}(2016){Shahbaz}, {Russell}, {Covino}, {Mooley}, {Fender}, \& {Rumsey}}]{shahbaz16}
{Shahbaz}, T., {Russell}, D.~M., {Covino}, S., {et~al.} 2016, \bibinfo{title}{{Evidence for magnetic field compression in shocks within the jet of V404 Cyg},} \mnras, 463, 1822, \dodoi{10.1093/mnras/stw2171}

\bibitem[{N.~I. {Shakura} \& R.~A. {Sunyaev}(1973){Shakura} \& {Sunyaev}}]{shakura73}
{Shakura}, N.~I., \& {Sunyaev}, R.~A. 1973, \bibinfo{title}{{Reprint of 1973A\&A....24..337S. Black holes in binary systems. Observational appearance.},} \aap, 500, 33

\bibitem[{S. {Sharma}(2017){Sharma}}]{Sharma2017}
{Sharma}, S. 2017, \bibinfo{title}{{Markov Chain Monte Carlo Methods for Bayesian Data Analysis in Astronomy},} \araa, 55, 213, \dodoi{10.1146/annurev-astro-082214-122339}

\bibitem[{M. {Shidatsu} {et~al.}(2017){Shidatsu}, {Tachibana}, {Yoshii}, {Negoro}, {Kawamuro}, {Iwakiri}, {Nakahira}, {Makishima}, {Ueda}, {Kawai}, {Serino}, \& {Kennea}}]{shidatsu17}
{Shidatsu}, M., {Tachibana}, Y., {Yoshii}, T., {et~al.} 2017, \bibinfo{title}{{Discovery of the New X-Ray Transient MAXI J1807+132: A Candidate of a Neutron Star Low-mass X-Ray Binary},} \apj, 850, 155, \dodoi{10.3847/1538-4357/aa93f0}

\bibitem[{M.~F. {Skrutskie} {et~al.}(2006){Skrutskie}, {Cutri}, {Stiening}, {Weinberg}, {Schneider}, {Carpenter}, {Beichman}, {Capps}, {Chester}, {Elias}, {Huchra}, {Liebert}, {Lonsdale}, {Monet}, {Price}, {Seitzer}, {Jarrett}, {Kirkpatrick}, {Gizis}, {Howard}, {Evans}, {Fowler}, {Fullmer}, {Hurt}, {Light}, {Kopan}, {Marsh}, {McCallon}, {Tam}, {Van Dyk}, \& {Wheelock}}]{skrutskie06}
{Skrutskie}, M.~F., {Cutri}, R.~M., {Stiening}, R., {et~al.} 2006, \bibinfo{title}{{The Two Micron All Sky Survey (2MASS)},} \aj, 131, 1163, \dodoi{10.1086/498708}

\bibitem[{J. {Smak}(1971){Smak}}]{smak71}
{Smak}, J. 1971, \bibinfo{title}{{Eruptive Binaries. II. U Geminorum},} \actaa, 21, 15

\bibitem[{J. {Smak}(1984){Smak}}]{smak84}
{Smak}, J. 1984, \bibinfo{title}{{Outbursts of dwarf novae.},} \pasp, 96, 5, \dodoi{10.1086/131295}

\bibitem[{R.~E. {Spencer} {et~al.}(2013){Spencer}, {Rushton}, {Ba{\l}uci{\'n}ska-Church}, {Paragi}, {Schulz}, {Wilms}, {Pooley}, \& {Church}}]{spencer13_AKH}
{Spencer}, R.~E., {Rushton}, A.~P., {Ba{\l}uci{\'n}ska-Church}, M., {et~al.} 2013, \bibinfo{title}{{Radio and X-ray observations of jet ejection in Cygnus X-2},} \mnras, 435, L48, \dodoi{10.1093/mnrasl/slt090}

\bibitem[{P.~B. {Stetson}(1987){Stetson}}]{Stetson1987}
{Stetson}, P.~B. 1987, \bibinfo{title}{{DAOPHOT - A computer program for crowded-field stellar photometry},} \pasp, 99, 191, \dodoi{10.1086/131977}

\bibitem[{P.~B. {Stetson}(1990){Stetson}}]{stetson90}
{Stetson}, P.~B. 1990, \bibinfo{title}{{On the Growth-Curve Method for Calibrating Stellar Photometry with CCDs},} \pasp, 102, 932, \dodoi{10.1086/132719}

\bibitem[{R.~A. {Sunyaev} \& L.~G. {Titarchuk}(1985){Sunyaev} \& {Titarchuk}}]{sunyaev85}
{Sunyaev}, R.~A., \& {Titarchuk}, L.~G. 1985, \bibinfo{title}{{Comptonization of low-frequency radiation in accretion disks Angular distribution and polarization of hard radiation},} \aap, 143, 374

\bibitem[{J. {Tan} {et~al.}(1992){Tan}, {Lewin}, {Hjellming}, {Penninx}, {van Paradijs}, {van der Klis}, \& {Mitsuda}}]{tan92}
{Tan}, J., {Lewin}, W.~H.~G., {Hjellming}, R.~M., {et~al.} 1992, \bibinfo{title}{{Simultaneous X-Ray and Radio Observations of GX 5-1},} \apj, 385, 314, \dodoi{10.1086/170940}

\bibitem[{A.~J. {Tetarenko} {et~al.}(2017){Tetarenko}, {Sivakoff}, {Miller-Jones}, {Rosolowsky}, {Petitpas}, {Gurwell}, {Wouterloot}, {Fender}, {Heinz}, {Maitra}, {Markoff}, {Migliari}, {Rupen}, {Rushton}, {Russell}, {Russell}, \& {Sarazin}}]{atetarenko17_AKH}
{Tetarenko}, A.~J., {Sivakoff}, G.~R., {Miller-Jones}, J.~C.~A., {et~al.} 2017, \bibinfo{title}{{Extreme jet ejections from the black hole X-ray binary V404 Cygni},} \mnras, 469, 3141, \dodoi{10.1093/mnras/stx1048}

\bibitem[{J.~L. {Tonry} {et~al.}(2018){Tonry}, {Denneau}, {Flewelling}, {Heinze}, {Onken}, {Smartt}, {Stalder}, {Weiland}, \& {Wolf}}]{tonry18}
{Tonry}, J.~L., {Denneau}, L., {Flewelling}, H., {et~al.} 2018, \bibinfo{title}{{The ATLAS All-Sky Stellar Reference Catalog},} \apj, 867, 105, \dodoi{10.3847/1538-4357/aae386}

\bibitem[{V. {Tudor} {et~al.}(2017){Tudor}, {Miller-Jones}, {Patruno}, {D'Angelo}, {Jonker}, {Russell}, {Russell}, {Bernardini}, {Lewis}, {Deller}, {Hessels}, {Migliari}, {Plotkin}, {Soria}, \& {Wijnands}}]{tudor17}
{Tudor}, V., {Miller-Jones}, J.~C.~A., {Patruno}, A., {et~al.} 2017, \bibinfo{title}{{Disc-jet coupling in low-luminosity accreting neutron stars},} \mnras, 470, 324, \dodoi{10.1093/mnras/stx1168}

\bibitem[{V. {Tudose} {et~al.}(2009){Tudose}, {Fender}, {Linares}, {Maitra}, \& {van der Klis}}]{tudose09}
{Tudose}, V., {Fender}, R.~P., {Linares}, M., {Maitra}, D., \& {van der Klis}, M. 2009, \bibinfo{title}{{The disc-jet coupling in the neutron star X-ray binary Aquila X-1},} \mnras, 400, 2111, \dodoi{10.1111/j.1365-2966.2009.15604.x}

\bibitem[{J. {van den Eijnden} {et~al.}(2021){van den Eijnden}, {Degenaar}, {Russell}, {Wijnands}, {Bahramian}, {Miller-Jones}, {Hern{\'a}ndez Santisteban}, {Gallo}, {Atri}, {Plotkin}, {Maccarone}, {Sivakoff}, {Miller}, {Reynolds}, {Russell}, {Maitra}, {Heinke}, {Armas Padilla}, \& {Shaw}}]{vandeneijnden21}
{van den Eijnden}, J., {Degenaar}, N., {Russell}, T.~D., {et~al.} 2021, \bibinfo{title}{{A new radio census of neutron star X-ray binaries},} \mnras, 507, 3899, \dodoi{10.1093/mnras/stab1995}

\bibitem[{J. {van Paradijs}(1996){van Paradijs}}]{vanparadijs96}
{van Paradijs}, J. 1996, \bibinfo{title}{{On the Accretion Instability in Soft X-Ray Transients},} \apjl, 464, L139, \dodoi{10.1086/310100}

\bibitem[{J. {van Paradijs} \& J.~E. {McClintock}(1994){van Paradijs} \& {McClintock}}]{vanparadijs94}
{van Paradijs}, J., \& {McClintock}, J.~E. 1994, \bibinfo{title}{{Absolute visual magnitudes of low-mass X-ray binaries.},} \aap, 290, 133

\bibitem[{J. {van Paradijs} \& F. {Verbunt}(1984){van Paradijs} \& {Verbunt}}]{vanparadijs84}
{van Paradijs}, J., \& {Verbunt}, F. 1984, in American Institute of Physics Conference Series, Vol. 115, High Energy Transients in AstroPhysics, ed. S.~E. {Woosley}, 49--62, \dodoi{10.1063/1.34556}

\bibitem[{A. {Veledina} {et~al.}(2013){Veledina}, {Poutanen}, \& {Vurm}}]{veledina13}
{Veledina}, A., {Poutanen}, J., \& {Vurm}, I. 2013, \bibinfo{title}{{Hot accretion flow in black hole binaries: a link connecting X-rays to the infrared},} \mnras, 430, 3196, \dodoi{10.1093/mnras/stt124}

\bibitem[{P. Virtanen {et~al.}(2020)Virtanen, Gommers, Oliphant, Haberland, Reddy, Cournapeau, Burovski, Peterson, Weckesser, Bright, {van der Walt}, Brett, Wilson, Millman, Mayorov, Nelson, Jones, Kern, Larson, Carey, Polat, Feng, Moore, {VanderPlas}, Laxalde, Perktold, Cimrman, Henriksen, Quintero, Harris, Archibald, Ribeiro, Pedregosa, {van Mulbregt}, \& {SciPy 1.0 Contributors}}]{scipy}
Virtanen, P., Gommers, R., Oliphant, T.~E., {et~al.} 2020, \bibinfo{title}{{{SciPy} 1.0: Fundamental Algorithms for Scientific Computing in Python},} Nature Methods, 17, 261, \dodoi{10.1038/s41592-019-0686-2}

\bibitem[{G.~B. {Zhang} {et~al.}(2019){Zhang}, {Bernardini}, {Russell}, {Gelfand}, {Lasota}, {Qasim}, {AlMannaei}, {Koljonen}, {Shaw}, {Lewis}, {Tomsick}, {Plotkin}, {Miller-Jones}, {Maitra}, {Homan}, {Charles}, {Kobel}, {Perez}, \& {Doran}}]{zhanggb19}
{Zhang}, G.~B., {Bernardini}, F., {Russell}, D.~M., {et~al.} 2019, \bibinfo{title}{{Bright Mini-outburst Ends the 12 yr Long Activity of the Black Hole Candidate Swift J1753.5-0127},} \apj, 876, 5, \dodoi{10.3847/1538-4357/ab12dd}

\bibitem[{C. {Zurita} {et~al.}(2006){Zurita}, {Torres}, {Steeghs}, {Rodr{\'\i}guez-Gil}, {Mu{\~n}oz-Darias}, {Casares}, {Shahbaz}, {Mart{\'\i}nez-Pais}, {Zhao}, {Garcia}, {Piccioni}, {Bartolini}, {Guarnieri}, {Bloom}, {Blake}, {Falco}, {Szentgyorgyi}, \& {Skrutskie}}]{zurita06}
{Zurita}, C., {Torres}, M.~A.~P., {Steeghs}, D., {et~al.} 2006, \bibinfo{title}{{The 2005 Outburst of the Halo Black Hole X-Ray Transient XTE J1118+480},} \apj, 644, 432, \dodoi{10.1086/503286}

\end{thebibliography}
\bibliographystyle{aasjournal}

\end{document}